\documentclass[usenatbib]{mn2e}

\usepackage[english]{babel}
\usepackage[latin1]{inputenc}
\usepackage[T1]{fontenc}

\usepackage{dcolumn}

\usepackage{amsmath}
\usepackage{arydshln}
\usepackage{bbold}
\usepackage{graphicx}
\usepackage{amsfonts}
\usepackage{amssymb}
\usepackage{hyperref}
\usepackage{latexsym}
\usepackage{color}

\newcommand{\be}{\begin{equation}}
\newcommand{\ee}{\end{equation}}
\newcommand{\bea}{\begin{eqnarray}}
\newcommand{\eea}{\end{eqnarray}}

\begin{document}


\title[Fourier analysis of multi-tracer cosmological surveys]{Fourier analysis of multi-tracer cosmological surveys}

\author[L.~R.~Abramo \& Lucas F. Secco \& Arthur Loureiro]
{L.~Raul Abramo\thanks{abramo@if.usp.br}, Lucas F. Secco and Arthur Loureiro
\\
Departamento de F\'{\i}sica Matem\'atica,
Instituto de F\'{\i}sica, Universidade de S\~{a}o Paulo, 
CP 66318, CEP 05314-970 S\~{a}o Paulo, Brazil}

\maketitle

\begin{abstract}
We present optimal quadratic estimators for the Fourier analysis 
of cosmological surveys that detect several different types of tracers 
of large-scale structure. 
Our estimators can be used to simultaneously fit the matter power 
spectrum and the biases of the tracers --- as well as
redshift-space distortions (RSDs), non-Gaussianities (NGs), or any other effects
that are manifested through differences between the clusterings of distinct species of tracers.
Our estimators reduce to the one by Feldman, Kaiser \& Peacock (ApJ 1994, FKP) 
in the case of a survey consisting of a single species of tracer.
We show that the multi-tracer estimators are unbiased, and that their 
covariance is given by the inverse of the multi-tracer Fisher matrix 
(Abramo, MNRAS 2013; Abramo \& Leonard, MNRAS 2013).
When the biases, RSDs and NGs are fixed to their fiducial values, and 
one is only interested in measuring the underlying power spectrum,  
our estimators are projected into the estimator found by Percival, Verde 
\& Peacock (MNRAS 2003). 
We have tested our estimators on simple (lognormal) simulated galaxy maps, 
and we show that it performs as expected, being equivalent or superior to 
the FKP method in all cases we analyzed. 
Finally, we have shown how to extend
the multi-tracer technique to include the 1-halo term of the power spectrum.
\end{abstract}



\section{Introduction}

Astrophysical surveys have come to occupy a central role in cosmology
\citep{York:2000gk,cole_2df_2005,Abbott:2005bi,scoville_cosmic_2007,adelman-mccarthy_sdss_2008,adelman-mccarthy_sixth_2008,PAN-STARRS,BOSS,2011MNRAS.415.2876B}.
Percent-level accuracies can now be reached on distance measurements at high redshifts,
both across and along the line-of-sight \citep{Andersonetal2012,Andersonetal2014}. 
This means more and better constraints on the
acceleration of the expansion rate of the Universe, on modified gravity \citep{LinderGrowth}, 
on non-Gaussian initial conditions \citep{NG2000,BartoloRev}, and about 
the role of massive neutrinos, among other applications.

The groundbreaking achievements of the Sloan Digital Sky Survey \citep{York:2000gk} 
are being surpassed by other surveys with higher completeness, wider wavelength coverage, and 
a larger range of redshifts \citep{BigBOSS,SUMIRE,PFS,LSST:2009pq}.
Some surveys will specialize in mapping very large volumes to an extremely high completeness, 
by employing imaging with narrow-band filters \citep{Benitez:2008fs,Benitez:2014}, or 
by resorting to low-resolution integral-field spectroscopy \citep{HETDEX}. 
In addition to galaxies, quasars can also serve both as sources of background light 
to investigate the intervening matter through the Ly-$\alpha$ forest \citep{Slosar2013}, 
or directly as tracers of the large-scale structure \citep{Croom2005,daAngela2005,Yahata2005,Shen2007,Ross2009,Sawangwit2012,2011arXiv1108.2657A,Leistedt2013,Leistedt2014}.
Another way in which
the 3D matter distribution can be mapped is through the 21cm hyperfine transition of neutral H,
and new radiotelescopes dedicated to measuring that line are being deployed or are in the planning stages 
\citep{CHIME,BINGO}. 

This intense activity points to an exciting future, where vast volumes of the Universe will be 
increasingly mapped in a variety of ways, and with different types of tracers of the large-scale structure.

However, these maps are not independent, since all tracers sit on the same distribution of 
dark matter. This points to a key obstacle on the way to explore the full power of 
these overlapping surveys: cosmic variance -- a particular case of sample variance,
where the sample is the set of modes of the density perturbations 
which were realized in some region of the Universe
from a (nearly) Gaussian random process.

Despite this fundamental limitation, it was pointed out by \citet{SeljakNG,McDonald:2008sh} 
that the bounds imposed by cosmic variance do not apply for many key physical 
quantities of interest.
In particular, by comparing the clustering of tracers with different biases one can measure 
some parameters to an accuracy which is basically unconstrained by cosmic variance.
This applies not only to bias itself, but to the matter growth rate of the RSDs, to the 
NG parameters $f_{NL}$ and $g_{NL}$, as well as any parameters that affect 
the relative clusterings of different tracers.
The consequences of this extraordinary windfall were explored in many papers -- see, e.g., \citet{Slosar09,GM2010,CaiCV,CaiWL,Hamaus2011,SDM,Hamaus2012,AL13,GAMA,Bull,Ferramacho,Hamaus14}.
It is important to stress that this additional information comes from measuring 
the two-point functions of the different tracers, as opposed to enhancing the signal-to-noise
by employing different statistics such as mass-weighting
\citep{Seljaketal09,2011MNRAS.412..995C,SmithMarian14}.

In \citet{AL13} we showed that the enhanced constraints of multi-tracer cosmological surveys 
are a straightforward consequence of the multi-tracer Fisher information matrix \citep{AbramoFisher}.
In the presence of $N$ different types of tracers, each one with a different bias,
there is a simple choice of variables which diagonalizes the multi-tracer Fisher matrix:
in addition to the underlying power spectrum (which is subject to the cosmic variance bounds), 
there are $N-1$ variables which correspond to the relative clustering strengths between the tracers.
These relative clustering strengths are not affected by cosmic variance,
and their measurements can be arbitrarily accurate, even if the survey has a finite volume.
In the case of a single tracer, the multi-tracer Fisher matrix reduces to the usual case treated in
the seminal paper by \citet{1994ApJ...426...23F} (henceforth FKP).

In this paper we use the multi-tracer Fisher matrix to derive optimal estimators for the 
redshift-space power spectra of an arbitrary number of different types of tracers of 
large-scale structure (see Sec. \ref{S:Estimator}). 
These tracers may overlap in some regions but not others, or not at all.
The tracers can be galaxies of different types, quasars, Ly-$\alpha$ absorption systems, etc.
One may also choose to trade individual objects by halos of different masses --- in which
case the bias of the tracers become the halo bias 
\citep{Seljaketal09,Hamaus2010,2011MNRAS.412..995C,SmithMarian14,SmithMarian15}.

Our formulas can be used in any of those situations, in real or in redshift-space, 
including effects such as scale-dependent bias.
An important cross-check is that a particular combination (or projection) of 
our estimators leads back to the estimator of \citet{Percival:2003pi} (henceforth PVP), 
--- namely, the PVP estimator 
follows from ours if the biases of the tracers, as well as the RSDs, are fixed to their 
true values, and one then computes the underlying matter power spectrum using the 
aggregated clustering information from all the tracers.

We have also incorporated the contribution of the 1-halo term to the covariance 
of the galaxy counts (see Sec. \ref{S:1-halo}) into the multi-tracer Fisher matrix.
In principle, the full covariance can be 
computed using the Halo Model \citep{CooraySheth}, together with appropriate 
halo occupation distributions for the tracers. 
Recently, \citet{SmithMarian15} presented an optimal estimator for the
power spectrum including all Halo Model corrections to the spectrum,
bispectrum and trispectrum. 
However, the \citet{SmithMarian15} estimator generalizes the estimator of 
\citet{Percival:2003pi}, while we have obtained estimators not only for the
matter power spectrum, but also for the biases, RSDs, NGs, etc.
Hence, we are only able to include the simplest correction to galaxy 
clustering from the halo model (the 1-halo term), but our framework allows
the simultaneous estimation of bias, RSDs and the power spectrum,
while \citet{SmithMarian15}, include all the corrections that can be computed on the basis of
the Halo Model, but their estimator applies only to the power spectrum, and assumes
prior knowledge of the bias of all the species, as well as of the shape of the RSDs.

The Fisher matrix and the covariance of the counts of the tracers
are the basic objects used to construct our estimators -- in fact, the 
optimal estimators are a type of Wiener filtering, in the sense that we are
basically weighting the data by the inverse of their covariance. 
We employ results and notation introduced in
\citet{AbramoFisher,AL13}, and the construction of the estimators 
follows the steps outlined in
\citet{Tegmark_Surveys_1997,1998ApJ...499..555T}.

The estimators were tested using simple mock galaxy catalogs based
on lognormal maps (Sec. \ref{S:Applications}).
We find that the empirical covariance of the 
power spectra is well approximated by the theoretical covariance (the
inverse of the multi-tracer Fisher matrix), confirming the optimality of the estimator.

The main formulas of this paper are derived in Sec. 3, and a practical
algorithm for the Fourier analysis of multi-tracer surveys 
is summarized in Sec. \ref{SubSec:Algo}.

This paper is organized as follows: in Sec. \ref{S:Information} we review the Fisher 
information matrix for single-tracer and multi-tracer cosmological surveys. In Sec. 
\ref{S:Estimator} we construct the optimal quadratic estimators on the basis of the 
covariance matrix for the data.
In Sec. \ref{Sec:Prop} we discuss the relationships between the multi-tracer estimators
and other methods, such as FKP and PVP, as well as the main features of the 
multi-tracer technique.
In Sec. \ref{S:Applications} we test the estimators in simple simulated maps, showing
that the empirical covariance matches closely the theoretical covariance --- 
which establishes that the multi-tracer estimators are unbiased and optimal.
In Sec. \ref{S:1-halo} we show how to include the 1-halo term in the estimators of the
multi-tracer power spectra. We also show there how to construct estimators for the 
1-halo term, and how to generalize the procedure to estimate simultaneously the 
2-halo and the 1-halo term of the power spectrum. 
We conclude in Section \ref{S:Conclusions}.


\section{The information in galaxy surveys}
\label{S:Information}

The matter power spectrum is defined through the expectation value 
$\langle \delta_m (\vec{k},z) \delta_m^*(\vec{k}',z) \rangle = (2\pi)^3 P_m(k,z) 
\delta_D (\vec{k}-\vec{k}')$, where $\delta_m(\vec{k})$ is the matter density contrast, and 
$\delta_D$ is the Dirac delta function.
However, galaxy surveys actually measure counts of tracers of the large-scale structure 
(galaxies and other extragalactic objects) in redshift space.
It is from those observable that we can then measure derived quantities such as 
the baryon acoustic oscillations \citep{eisenstein_cosmic_1998,blake_probing_2003,
seo_probing_2003}, or the pattern of redshift-space distortions 
\citep{Kaiser87,HamiltonRev05a,HamiltonRev05b}.

For a tracer of type $\alpha$, whose counts as a function of (redshift-space) position are
$n_\alpha(\vec{x})$, the density contrast is 
$\delta_\alpha (\vec{x}) = n_\alpha(\vec{x})/\bar{n}_\alpha(\vec{x}) -1$.
The mean number densities $\bar{n}_\alpha$ should
reflect the spatial modulations of the observed numbers of galaxies which are
due to the instrument, the strategy and schedule of observations, as well as any other
factors unrelated to the redshift-space cosmological fluctuations.

If we assume that bias is linear and deterministic, then
in the distant-observer approximation the redshift-space fluctuations in the counts of 
the $\alpha$-type galaxies are
related to the underlying mass fluctuations by the relation
$ \delta_\alpha (\vec{k},z) \simeq [b_\alpha + f \, \mu_k^2] \delta_m(\vec{k},z) \, $.
Here $b_\alpha$ is the bias of the tracer species $\alpha$,
$f(z)$ is the matter growth rate, and $\mu_k=\hat{k}\cdot\hat{r}$ 
is the cosine of the angle between the Fourier mode and the line of sight.

The index $\alpha$ (we employ greek letters to denote different tracer species) can be
any kind of discriminant of the types of tracers of large-scale structure:
it may stand for luminosity, morphological type, star formation rate, equivalent width of some
emission line, or a combination of those.
One may also regard the dark matter halos themselves as the tracers, in which case 
$\alpha$ would stand for the halo mass (or some proxy for it, such as richness), 
and the bias $b_\alpha$ becomes halo bias.

There are some complicating factors in this description.
First, structure formation should introduce a scale dependence for the bias, 
as well as some degree of stochasticity \citep{Benson:1999mva,BiasLahav,BiasWeinberg,BiasRavi}. 
Second, the initial conditions may contain non-Gaussian features,
which would manifest themselves as an additional scale-dependent bias
\citep{BartoloRev,Sefusatti2007,NGDalal}. 
Third, the velocity dispersion from random motions inside halos will smear the galaxy
density contrast, affecting the shape of RSDs.
In fact, the RSD parameters and angular dependence can inherit scale-dependent 
non-linear corrections \citep{Raccanelli2012}

Hence, in practice it is more useful to regard ``bias'' 
as a more general function of redshift, scale, and angle with the line-of-sight
that should be determined by observations, 
and {\em define} the clustering of a species $\alpha$ as:
\be
P_\alpha \equiv B_\alpha^2  \, P_m \; ,
\ee
where $B_\alpha = b_\alpha + f\mu_k^2 + \Delta b_{NG}$ is an {\em effective bias}.
This effective bias can include not only RSDs and non-Gaussianities (NGs), 
but also scale-dependence of the bias, or any other effect that distorts the 
power spectrum of the tracers relative to the underlying matter distribution.

In principle, everything depends on $\vec{x}$ and $\vec{k}$,
but one can regard $x$ (i.e., the radial position) as standing in for 
redshift, so we could also write $B_\alpha = B_\alpha (z,k,\mu_k)$, and 
$P_m = P_m (z,k)$.
Since the matter power spectrum, bias, RSDs, as well as NGs and other corrections, 
are just subproducts of clustering measurements for 
all the available tracers in a 
given survey, the problem we must address is how one can optimally estimate 
the clusterings $P_\alpha (\vec{x},\vec{k})$. Our approach also means that cross-correlations 
are expressed as $P_{\alpha \beta} = B_\alpha B_\beta P_m$.


\subsection{Optimal estimators and the Fisher matrix}

The Fisher information matrix can be constructed from the data covariance
after a series of simple steps -- for a review in the
specific context of cosmological surveys, see, e.g., 
\citet{1997ApJ...480...22T,Tegmark_Surveys_1997,1998ApJ...499..555T}.

Let's assume that the measured quantities (the data) are $d_i$, such that, 
for simplicity, their expectation values vanish: $\langle d_i \rangle = 0$. 
The data covariance is then
$C_{ij} = Cov(d_i,d_j) = \langle d_i d_j \rangle $. 
From this data we would like to extract some set of parameters $p_\mu$, 
whose likelihoods we assume to be approximately described by a 
multivariate Gaussian. Under these conditions, 
the Fisher information matrix is given by:
\bea
\label{Def:F}
F_{\mu \nu} &\equiv& F[p_\mu,p_\nu] 
= - \langle \frac{\partial^2 \log {\cal{L}}}{\partial p_\mu \partial p_\nu}  \rangle
\\ \nonumber
&\simeq&  \frac12 \sum_{ijkl}  C_{ij}^{-1} \frac{\partial C_{jk}}{\partial p_\mu} C_{kl}^{-1} \frac{\partial C_{li}}{\partial p_\nu}  \; ,
\\ \nonumber
&=& \frac12 \, {\rm Tr} \left[ C^{-1} \frac{\partial C}{\partial p_\mu} C^{-1} \frac{\partial C}{\partial p_\nu} \right] \; ,
\eea
where the second line follows from the assumption of near-Gaussianity of the 
likelihood near its maximum.

Now suppose that estimators $\hat{p}_\mu$ can be constructed, such that their covariance
$Cov(\hat{p}_\mu,\hat{p}_\nu) = F^{-1}_{\mu\nu}$. These estimators must then be optimal, 
in the sense that they saturate the Cram\'er-Rao bound: 
$Cov(\hat{p}_\mu,\hat{p}_\nu) \geq F^{-1}_{\mu\nu}$.

There are in fact such estimators, which can be constructed after a 
few simple steps, and which employ the same basic 
objects that appear in the Fisher matrix.
The first step is to create the quadratic form:
\be
\label{Def:qhat}
\hat{q}_\mu \equiv \sum_{ij} E_\mu^{ij} d_i d_j - \Delta_\mu \; ,
\ee
where 
\be
\label{Def:E}
E_\mu^{ij} = \frac12 \sum_{kl} C^{-1}_{ik}  \frac{\partial C_{kl} }{\partial p_\mu} C^{-1}_{lj} .
\ee
Here, $\Delta_\mu$ serves to subtract any possible bias the estimators may
have, such that we end up with unbiased estimators whose expectation values
coincide with their fiducial values.

Gaussianity of the data (a key underlying assumption) implies that the 4-point function
$\langle d_i d_j d_k d_l \rangle = C_{ij} C_{kl} + C_{ik} C_{jl} + C_{il} C_{jk}$, and from it follows, after some algebra, that the covariance of the quadratic form above is 
$Cov(\hat{q}_\mu,\hat{q}_\nu) = F_{\mu\nu}$.

The final step is to define the optimal quadratic estimators in such a way that their 
covariance is the {\em inverse} of the Fisher matrix. Clearly, then, the estimators:
\be
\label{Def:phat}
\hat{p}_\mu = \sum_\alpha F^{-1}_{\mu \alpha} \, \hat{q}_\alpha \; ,
\ee
satisfy that condition. Finally, with the definition:
\bea
\label{Eq:Dbias}
\Delta_\alpha &\equiv&  \sum_{ij} E_\alpha^{ij} C_{ij} - \sum_\beta F_{\alpha \beta} \, 
\bar{p}_\beta 
\\ \nonumber 
& = & 
\frac12
\sum_{ki} C_{ik}^{-1} \frac{\partial C_{ki} }{\partial p_\mu}   - 
\sum_\beta F_{\alpha \beta} \, \bar{p}_\beta 
\; ,
\eea
we obtain estimators which are also {\em unbiased}: their expectation values are 
equal to the fiducial values of those parameters: $\langle \hat{p}_\mu \rangle \to \bar{p}_\mu$.


\subsection{Single-species Fisher matrix}

The most basic sources of uncertainty in galaxy surveys are cosmic variance and shot noise.
In cosmological surveys which target a single species of tracer,
the optimal estimator for the galaxy power spectrum was derived 
by FKP \citep{1994ApJ...426...23F}. 
The corresponding Fisher information matrix
was derived by \citet{Tegmark_Surveys_1997,1998ApJ...499..555T}, who
also showed that the FKP estimator follows from the construction presented in the
previous Section.

The FKP Fisher matrix for a survey of some galaxy type $\alpha$ can be written as:
\be
\label{Eq:FKP_Fisher}
F[\theta_i , \theta_j ] = 
\int \frac{d^3 x \, d^3 k}{(2\pi)^3}
\, \frac{\partial \log P_\alpha}{\partial \theta^i} 
\, {\cal{F}}_\alpha
\, \frac{\partial \log P_\alpha}{\partial \theta^j} \; ,
\ee
where the Fisher {\em information density} in phase space associated 
with the tracer $\alpha$ is:
\be
\label{Eq:F_alpha}
{\cal{F}}_\alpha (\vec{x},\vec{k}) = \frac12 \left( \frac{{\cal{P}}_\alpha}{1+{\cal{P}}_\alpha} \right)^2 \; .
\ee
In the expression above we have defined a dimensionless ``clustering strength'' 
of the tracer $\alpha$, which is just the power spectrum in units of (Poissonian) shot noise: 
\be
\label{Eq:Palpha}
{\cal{P}}_\alpha(\vec{x},\vec{k}) \, \equiv \,  \bar{n}_\alpha(\vec{x}) P_\alpha (\vec{x},\vec{k}\,) \; .
\ee
In the limit of arbitrarily high clustering strength ($1/\bar{n}_\alpha \to 0$, ${\cal{P}}_\alpha \to \infty$),
the Fisher information density saturates the limit ${\cal{F}}_\alpha \rightarrow \frac12$. 
Hence, for a survey of a single species of tracer there is an upper limit to 
the information which can be extracted 
from a finite volume and from a finite range of Fourier modes.
This is nothing but a restatement of the limits imposed by cosmic variance.

At this point it is useful to recall how, in practice, one can extract limits on the amplitude of
the power spectrum out of the Fisher matrix. In that case, the parameters of the Fisher matrix 
are the values of the matter power spectrum at given bandpowers (i.e., at
some bins in Fourier space), obtained from a given survey volume.
Consider, then, the parameters:
\bea
\label{Eq:parP}
\theta^i &\rightarrow& P_{\alpha,i} \equiv \langle P_\alpha(\vec{x},\vec{k}) \rangle_i 
\\ \nonumber
&=& \frac{1}{V_{\vec{x}_i}}  \int_{V_{\vec{x}_i}} d^3 x \;
\frac{1}{V_{\vec{k}_i}} \int_{V_{\vec{k}_i}} \frac{d^3 k }{(2\pi)^3} \;
P_\alpha(\vec{x},\vec{k})
\; ,
\eea
where $\vec{x}_i$ represents a bin in real space (e.g., a redshift slice $z_i$) with volume 
$V_{\vec{x}_i}$, and $\vec{k}_i$ represents a bin in Fourier space 
(i.e., a bandpower $k_i$, and an angular bin $\mu_{k,i}$)
of volume $V_{\vec{k}_i}$.
In that case we should compute the Jacobian 
$\partial P_\alpha(\vec{x},\vec{k})/\partial P_{\alpha,i}$ inside Eq. (\ref{Eq:FKP_Fisher}).
It is useful to regard such an object in terms of functional derivatives
\footnote{In fact, all partial derivatives used in connection with the Fisher
matrix should be replaced by functional derivatives in the continuum limit. 
It is only when we use bins (in real space and/or Fourier space) that these
functional derivatives are converted to partial derivatives. Nevertheless, in order to keep 
the notation as simple as possible, we employ the same notation for both.}.  
Using:
\be
\label{Eq:funcder}
\frac{\partial f(\vec{x},\vec{k})}{\partial f(\vec{x}{\,}',\vec{k}')} = (2\pi)^3 \delta_D(\vec{x}-\vec{x}{\,}') \delta_D(\vec{k}-\vec{k}') \; ,
\ee
one can easily derive that the inverse of the Jacobian is:
\be
\label{Eq:dPi}
\frac{\partial P_{\alpha,i}}{\partial P_\alpha(\vec{x},\vec{k})} = \frac{1}{V_{\vec{x}_i} \, V_{\vec{k}_i}} \; .
\ee
Therefore, the Jacobian $\partial P_\alpha(\vec{x},\vec{k})/\partial P_{\alpha,i}$ has the effect of
limiting integrations in phase space, $\int d^3 x \, d^3 k/(2\pi)^3 [ \cdots ]$, 
to the phase space volume of the bin $i$.
Since this type of object will reappear later on, we employ the notation 
$\delta^i_{\vec{x},\vec{k}}$ to express the restriction of a phase space integral to a
certain volume $V_i$, and we use the same notation to indicate 
restrictions in integrals over position space, $\delta^i_{\vec{x}}$, or 
Fourier space, $\delta^i_{\vec{k}}$. Hence, according to this notation:
\be
\int \frac{d^3 x \, d^3 k}{(2\pi)^3}  [\cdots] \times \delta^i_{\vec{x},\vec{k}} =
\int_{V_i} \frac{d^3 x  \, d^3 k}{(2\pi)^3}  [\cdots] \; .
\ee
Moreover, for non-overlapping bins $i$ and $j$ it follows that:
\be
\int \frac{d^3 x  \, d^3 k}{(2\pi)^3}  [\cdots] \times \delta^i_{\vec{x},\vec{k}} \times \delta^j_{\vec{x},\vec{k}} =
\delta_{ij} \, \int_{V_i} \frac{d^3 x  \, d^3 k}{(2\pi)^3}  [\cdots] \; .
\ee

With these identities in mind, it is trivial to see that when
using $P_{\alpha,i}$ as parameters, Eq. (\ref{Eq:FKP_Fisher}) reduces to:
\be
\label{Eq:FKP_P}
F_{\alpha  ,  i  j} = F[P_{\alpha, i}, P_{\alpha, j} ]  = \frac{\delta_{i j}}{P_{\alpha,i}^2}  \, 
\int_{V_i} \frac{d^3 x  \, d^3 k}{(2\pi)^3}  
\, {\cal{F}}_{\alpha}
\; .
\ee
A more familiar form for this equation follows if we revert to the definition of
averages over real- and Fourier-space bins:
\be
\label{Eq:FKP_P}
F_{\alpha  ,  i  j}  =
\frac{ \delta_{i j} \, V_{\vec{k}_i} }{P_{\alpha,i}^2} \,
\int_{V_{\vec{x}_i}} d^3 x \, 
\left\langle {\cal{F}}_{\alpha} \right\rangle_{\vec{k}_i}
\; .
\ee
Up to a factor of 2, the integral over position space in the equation above defines the 
usual {\em effective volume} 
\citep{Tegmark_Surveys_1997,1998ApJ...499..555T}. The uncertainty in the amplitude of 
the power spectrum at the bin $i$ is therefore given by the covariance 
$Cov[P_{\alpha,i},P_{\alpha,j}] = F^{-1}_{\alpha  ,  i  j}$, which 
is diagonal in the Fourier modes --- see, however, \citet{AbramoFisher}.
The relative uncertainty in
the bandpowers of the power spectrum is then given by the well-known expression:
\be
\label{SigmaFKP}
\frac{\sigma^2_{P_{\alpha,i}}}{P_{\alpha,i}^2} = 
\frac{1}{V_i  \, \langle {\cal{F}}_{\alpha} \rangle_i } \; ,
\ee
where $V_i = V_{\vec{x}_i} \, V_{\vec{k}_i}$ is the phase space volume of the
bin $i$, and $\langle \cdots \rangle_i$ denotes an average over the phase space bin.
When the number density of the tracer is very high, ${\cal{P}}_{\alpha} \gg 1$ and
${\cal{F}}_{\alpha} \to 1/2$, and if that is the case, then the survey is dominated by cosmic 
variance, $\sigma_{P_{\alpha,i}}/{P_{\alpha,i}} \to \sqrt{2/ V_i}$. The phase space
volume gives the number of modes of the bin $k_i$ that fit in the physical volume 
$V_{\vec{x}_i}$, and the factor
of 2 comes from the fact that the density contrast is real.


\subsection{Multi-tracer Fisher matrix}

Galaxy surveys can detect a wide variety of objects: galaxies of different types, quasars, 
Ly-$\alpha$ emmitters, Ly-$\alpha$ absorbers, etc. In the future all this data will coalesce
into multi-layer maps of the observable Universe, containting many different kinds of 
objects which can be regarded as tracers of the large-scale structure.

The multi-tracer Fisher information matrix describes how
the contributions of cosmic variance and shot noise affect the signal-to-noise ratio 
(SNR) of the
observables we are trying to measure -- namely, the clustering properties of the tracers.
While the nature of shot noise remains basically the same in the presence of 
multiple tracers, the effect of cosmic variance, which is shared among all tracers, mixes the 
different components.

The first authors to write a multi-tracer Fisher matrix (or, equivalently, a covariance matrix
for the power spectra) were
\citet{White:2008jy}, \citet{McDonald:2008sh} and \citet{Hamaus2012}.
In \citet{AbramoFisher}, we derived the multi-tracer Fisher directly from 
the covariance of the counts of the tracers (the ``pixel covariance''). 
The basic difference between  the approaches of 
\citet{White:2008jy,McDonald:2008sh} and ours is that those authors regard the cross-power 
spectra as independent parameters, while
we implicitly assume that, for the purposes of estimating the power spectra from the data, 
the cross-spectra are determined by the auto-spectra 
-- see, however, \citet{Swansonetal08} and \citet{Bonoli} for situations 
where this may not be true.
The Fisher matrix computed in Eq. (21) of \citet{Hamaus2012} also reduces to ours, if 
the cross-correlations are unaffected by shot noise -- see also \citet{Smith09,SmithMarian14,SmithMarian15}.

We now show how to obtain the multi-tracer Fisher matrix from first principles.
The generalization of Eq. (\ref{Def:F}) in the present context is:
\bea
\nonumber
F_{\mu, i \,; \, \nu, j} &=& 
\sum_{\alpha \beta \gamma \sigma}  
\int d^3 x \, d^3 x' \, d^3 x'' \, d^3 x''' 
\\ \nonumber
&\times&
C_{\alpha \beta}^{-1} (\vec{x},\vec{x}{ \, }') 
\frac{\partial C_{\beta \gamma} (\vec{x}{ \, }',\vec{x}{ \, }'')}{\partial P_{\mu,i}} 
\\ \label{Def:FC}
&\times&
C_{\gamma \sigma}^{-1} (\vec{x}{ \, }'',\vec{x}{ \, }''') 
\frac{\partial C_{\sigma \alpha} (\vec{x}{ \, }''',\vec{x})}{\partial P_{\nu,j}} 
 \; .
\eea

Let's express the covariance of tracer counts as:
\bea
\label{Eq:CovCounts}
C_{\alpha\beta}(\vec{x},\vec{x}{ \, }') &=& \xi_{\alpha\beta} (\vec{x},\vec{x}{ \, }') +
\frac{\delta_{\alpha\beta}}{\bar{n}_\alpha (\vec{x})} \, \delta_D(\vec{x}-\vec{x}{ \, }')
\\ \nonumber
&=& \int \frac{d^3 k}{(2\pi)^3} e^{i \vec{k} \cdot (\vec{x}-\vec{x}{ \, }')} 
\\ \nonumber
& & \times \left[ B_\alpha(\vec{x},\vec{k}) P_m (\bar{x},\vec{k}) B_\beta(\vec{x}{ \, }',\vec{k}) 
+ \frac{\delta_{\alpha\beta}}{\bar{n}_\alpha (\vec{x})} \right]
\; ,
\eea
where $\bar{x}$ denotes the mean position (or redshift) in which the matter power
spectrum is evaluated.
A key difficulty with the covariance of counts is
that, in any realistic situation, it cannot be inverted. However, 
if the effective biases and the power spectrum depend weakly on $\vec{k}$, then
it is a fair approximation to integrate the complex exponential in Eq. (\ref{Eq:CovCounts}) 
into a Dirac delta-function, and to pull the rest of the integrand outside of the integral \citep{HamiltonRev05a,HamiltonRev05b}. This imply taking the approximation that:
\be
\label{Eq:CovLim}
C_{\alpha\beta}(\vec{x},\vec{x}{ \, }') \rightarrow 
\delta_D (\vec{x}-\vec{x}{ \, }') \, \times \left[ 
\frac{\delta_{\alpha\beta}}{\bar{n}_\alpha}  + B_\alpha \, P_m B_\beta \right]
\; .
\ee
This expression can now be easily inverted, as we will show next. 

The inverse of the covariance should obey the property:
\bea
\sum_\beta \int d^3 x' C_{\alpha\beta}^{-1} (\vec{x},\vec{x}{ \, }') C_{\beta\gamma} (\vec{x}{ \, }',\vec{x}{ \, }'')
&=& 
\\ \nonumber
\sum_\beta \int d^3 x' C_{\alpha\beta} (\vec{x},\vec{x}{ \, }') C_{\beta\gamma}^{-1} (\vec{x}{ \, }',\vec{x}{ \, }'')
&=&
\delta_{\alpha\gamma} \, \delta_D (\vec{x}-\vec{x}{ \, }'') \; .
\eea
Since $\int d^3 x' \delta_D (\vec{x}-\vec{x}{ \, }')  \delta_D (\vec{x}{ \, }'-\vec{x}{ \, }'') =  \delta_D (\vec{x}-\vec{x}{ \, }'')$,
all we have to do is to invert the matrix inside the square brackets in Eq. (\ref{Eq:CovLim}).
But matrices of the type $M_{\alpha\beta} = \delta_{\alpha\beta} + v_\alpha v_\beta$ 
can be easily inverted, in fact 
$M^{-1}_{\alpha\beta} = \delta_{\alpha\beta} - v_\alpha v_\beta/(1+v^2)$,
where $v^2 = \sum_\gamma v_\gamma^2$. A simple generalization of this 
simple case leads immediately to:
\be
\label{Eq:InvCov}
C_{\alpha\beta}^{-1}(\vec{x},\vec{x}{ \, }') \rightarrow 
\delta_D (\vec{x}-\vec{x}{ \, }') \, \times 
\left[ \delta_{\alpha\beta} \, \bar{n}_\alpha -
\bar{n}_\alpha \frac{B_\alpha \, P_m B_\beta }{1 + {\cal{P}}}  \bar{n}_\beta
\right]
\; ,
\ee
where we define the {\em total clustering strength} as the sum of all clustering strengths:
\be
\label{Eq:Ptot}
{\cal{P}} (\vec{x},\vec{k}) = 
\sum_\mu \, \bar{n}_\mu \, (\vec{x}) \, B_\mu^2  (\vec{x},\vec{k}) \, 
P_m (\vec{x},\vec{k}) 
= \sum_\mu \, {\cal{P}}_\mu \; .
\ee

The problem with Eqs. (\ref{Eq:CovLim}) and (\ref{Eq:InvCov})
is that they refer to a scale $\vec{k}$ which does not exist in the original expression, 
Eq. (\ref{Eq:CovCounts}). In fact, Eqs. (\ref{Eq:CovLim})-(\ref{Eq:InvCov}) treat
the positions of the two-point function, $\vec{x}$ and $\vec{x}{ \, }'$ as one and the same, 
due to the Dirac delta-function.
Hence, one should think of the Fourier mode $\vec{k}$ which is implicit 
in Eqs. (\ref{Eq:CovLim})-(\ref{Eq:InvCov}), as the reciprocal of some typical 
physical distance between $\vec{x}$ and $\vec{x}{ \, }'$, 
and in that sense, its role is to limit the scope of that distance in 
expressions involving these approximations. Notice that this issue appears 
already in the FKP and PVP methods, and we do not present any new 
development regarding this point.

Coming back to Eq. (\ref{Def:FC}), we see that the last step before  
constructing the Fisher 
matrix is the computation of the term
$\partial C_{\alpha\beta}(\vec{x},\vec{x}{ \, }') / \partial P_{\mu,i}$.
Once again, it is useful to employ the notion of functional derivatives 
and the results of the previous Section. 
Using the second line of Eq. (\ref{Eq:CovCounts}) and the fact that 
$\partial P_{\alpha}(\vec{x},\vec{k}) / \partial P_{\mu,i} = \delta^i_{\vec{x},\vec{k}}$,
we find, after some rearrangement, that:
\bea
\nonumber
\frac{\partial C_{\alpha\beta} (\vec{x},\vec{x}{ \, }')}{\partial P_{\mu,i}}
&=& \int \frac{d^3k }{(2\pi)^3} 
e^{i \vec{k} \cdot (\vec{x}-\vec{x}{ \, }')} \left( \delta_{\alpha\mu} \delta^i_{\vec{x},\vec{k}} +
\delta_{\beta\mu} \delta^i_{\vec{x}{ \, }',\vec{k}} \right)
\\ \label{Eq:dCdP}
& & 
\times \frac{B_\alpha (\vec{x},\vec{k}) B_\beta (\vec{x}{ \, }',\vec{k})}{2 \, B_{\mu}^2(\vec{x}_i,\vec{k}_i)} 
\; .
\eea
Notice that this object should be regarded as an operator: 
when it acts on functions of $\vec{k}$ it causes an integration over 
Fourier space, which is restricted to the volume of the bin $\vec{k}_i$ by
the presence of the $\delta^i_{\vec{x},\vec{k}}$ and $\delta^i_{\vec{x}{\,}',\vec{k}}$.
Apart from a volume factor $V_{\vec{k}_i}$ this integration is nothing but 
an average over the Fourier bin $\vec{k}_i$.

We can now obtain the Fisher matrix by substituting Eqs. (\ref{Eq:InvCov}) 
and (\ref{Eq:dCdP}) into Eq. (\ref{Def:FC}).
The result, after a bit of algebra, is that:
\be
\label{Eq:FBin}
F_{\mu,i \, ; \, \nu,j} = \frac{\delta_{ij}}{P_{\mu,i} P_{\nu,i}} 
\int_{V_i} \frac{d^3 x \, d^3 k}{(2\pi)^3} {\cal{F}}_{\mu\nu} \; ,
\ee
where:
\be
\label{Def:Fcurved}
{\cal{F}}_{\mu\nu} (\vec{x},\vec{k}) = \frac{1}{4} 
\frac{\delta_{\mu\nu} {\cal{P}}_\mu {\cal{P}} (1+{\cal{P}}) + {\cal{P}}_\mu {\cal{P}}_\nu (1-{\cal{P}})}{(1+{\cal{P}})^2} \; .
\ee
Eq. (\ref{Def:Fcurved}) is in fact the Fisher
information density per unit of phase space volume for 
$\log {\cal{P}}_\mu$  \citep{AbramoFisher}:
\bea
\nonumber
& & F[ \, \log {\cal{P}}_{\mu} (\vec{x},\vec{k}) \, , \, \log {\cal{P}}_{\nu}(\vec{x}{\,}',\vec{k}') \, ] 
\\ \label{Eq:DefCurveFcont}
 & & \quad \quad \quad \quad = (2\pi)^3 \delta_D(\vec{x}-\vec{x}{\,}') \delta_D(\vec{k}-\vec{k}{\,}') 
{\cal{F}}_{\mu\nu} (\vec{x},\vec{k}) \; ,
\eea
or, equivalently, in bins of finite volume:
\be
\label{Eq:DefCurveF}
F[ \, \log {\cal{P}}_{\mu;i}  \, , \, \log {\cal{P}}_{\nu;j} \, ] 
 = \delta_{ij} {\cal{F}}_{\mu\nu} (\vec{x}_i,\vec{k}_i) \; .
\ee
One can easily check that the multi-tracer Fisher matrix of Eq. (\ref{Eq:FBin}) 
reduces to the FKP Fisher matrix, Eq. (\ref{Eq:FKP_P}),
when there is only one type of tracer.


\section{The optimal multi-tracer quadratic estimators}
\label{S:Estimator}

Starting from the Fisher matrix of Eq. (\ref{Eq:FBin}), and with the help of 
Eqs. (\ref{Eq:InvCov}) and (\ref{Eq:dCdP}), we are in a position to implement 
the construction of the estimators which was presented in Section 2.1. 
For now we will not discuss the role of random maps,
which help to subtract spurious fluctuations that could be generated by 
modulations on the mean number of tracers, $\bar{n}_\mu$. The calculations below
are exactly the same with or without the random maps, so we come back to this
issue at the end of this Section, after we have shown how to construct the multi-tracer
estimators.

Since our data are the density contrasts of the tracers, the quadratic form 
of Eq. (\ref{Def:qhat}) becomes:
\be
\label{Def:Qhat}
\hat{Q}_{\mu,i} = \sum_{\alpha\beta} \int d^3 x \, d^3 x' \,
E^{\mu,i}_{\alpha \beta} (\vec{x},\vec{x}{\,}'; \vec{x}_i, \vec{k}_i) 
\delta_\alpha(\vec{x}) \,
\delta_\beta(\vec{x}{ \, }') 
-\delta Q_{\mu,i}
\; ,
\ee
where $\delta Q_{\mu,i}$ ensures that the estimators are unbiased, and, 
according to the appropriate generalization of Eq. (\ref{Def:E}):
\be
\label{Def:Emu}
E^{\mu,i}_{\alpha \beta} = \frac12
\sum_{\sigma \gamma} \int d^3 y \, d^3 y' 
C_{\alpha\sigma}^{-1} (\vec{x},\vec{y}) 
\frac{\partial C_{\sigma\gamma} (\vec{y},\vec{y}{\,}')}{\partial P_{\mu,i}}
C_{\gamma\beta}^{-1} (\vec{y}{ \, }',\vec{x}{ \, }')  \; .
\ee
Inserting Eqs. (\ref{Eq:InvCov}) and (\ref{Eq:dCdP}) into Eq. (\ref{Def:Emu}), and
then back on Eq. (\ref{Def:Qhat}), leads to
the following expression for the quadratic form:
\bea
\label{Eq:Qhat}
\hat{Q}_{\mu,i} &=&
\frac{1}{4 \, B_{\mu,i}^2 } 
\sum_{\sigma \gamma}
\int d^3 x \, d^3 x' \int \frac{d^3 k}{(2\pi)^3}
e^{i \vec{k} \cdot (\vec{x} - \vec{x}{ \, }')}
\\ 
\nonumber
& & \times 
f_\sigma(\vec{x},\vec{k}) 
\left( 
\delta_{\sigma\mu} \delta^i_{\vec{x},\vec{k}} + 
\delta_{\gamma\mu} \delta^i_{\vec{x}{ }',\vec{k}}
\right) 
f_\gamma(\vec{x}{\,}',\vec{k}) - \delta Q_{\mu,i}\; ,
\eea
where
\be
\label{Def:f}
f_\sigma (\vec{x},\vec{k}) = \sum_\alpha w_{\sigma\alpha}(\vec{x},\vec{k}) 
\, \delta_\alpha(\vec{x}) \; 
\ee
are ``weighted density contrasts'' for the tracers. The weights are:
\be
\label{Def:w}
w_{\sigma\alpha} (\vec{x},\vec{k}) 
= \left[
\delta_{\sigma\alpha} - 
\frac{{\cal{P}}_\sigma (\vec{x},\vec{k})}{1+{\cal{P}}(\vec{x},\vec{k}) }
\right]
\bar{n}_\alpha B_\alpha (\vec{x},\vec{k}) \; .
\ee
These weights are the generalization of the FKP weights 
\citep{1994ApJ...426...23F} for the case of multiple tracers of large-scale structure. 
As we will prove in a moment, Eq. (\ref{Def:w}) defines the optimal weights for 
maps containing an arbitrary number of different types of tracers.
In the case of a single species of tracer, the weights for the density contrast reduce
to $w=\bar{n} B /(1+\bar{n} B^2 P_m)$, which is precisely the FKP weight for the 
density contrast -- except for a normalization, whose origin and purpose will 
become clearer soon.

Returning to Eq. (\ref{Eq:Qhat}), notice that the Kronecker delta functions are
accompanied by their respective restrictions over the phase space volume of the bin
where we are estimating the quantities of interest (in our case,
the $P_{\mu,i}$), hence:
\bea
\nonumber
\hat{Q}_{\mu,i} &=&
\frac{1}{4 \, B_{\mu,i}^2 } \int_{V_{\vec{k}_i}} \frac{d^3 k}{(2\pi)^3}
\\ 
\label{Eq:Qhat2}
& & \times 
\int_{V_{\vec{x}_i}} d^3 x \, e^{i \vec{k}_i \cdot \vec{x} } f_\mu(\vec{x})
\int d^3 x' \, e^{- i \vec{k}_i \cdot  \vec{x}{ \, }'} f (\vec{x} {\,} ') 
\\ \nonumber
& & + \; {\rm c.\, c.} \;  - \delta Q_{\mu,i} \; ,
\eea
where:
\be
\label{ftot}
f = \sum_\sigma f_\sigma = 
\frac{1}{1+{\cal{P}}} \sum_\sigma \bar{n}_\sigma B_\sigma \delta_\sigma \; .
\ee
Hence, the spatial integral over $f_\mu$ covers only the bin volume 
$V_{\vec{x}_i}$, while the spatial integral over $f$ should be performed over 
the whole volume of the survey. Although this is a subtlety which is present already in
the Fourier analys \`a la FKP, in practice we are always considering data on finite
volume bins, and all integrations are performed inside each one of those bins.
Nevertheless, a more rigorous treatment would dictate that one of the 
Fourier integrations be performed over the whole available volume of the survey,
while the other would be carried out over the volume of the particular 
bin under consideration.

In what follows we will ignore this subtlety, and will consider that both integrations
over spatial volume result in the Fourier transforms $\tilde{f}_\mu$ and $\tilde{f}$. 
We then obtain that:
\bea
\label{Eq:Qhat3}
\hat{Q}_{\mu,i} =
\frac{1}{4 \, B_{\mu,i}^2 } \int_{V_{\vec{k}_i}} \frac{d^3 k}{(2\pi)^3} \,
\left[ \tilde{f}_\mu(\vec{k}) \,
\tilde{f}^* (\vec{k})
+ \; {\rm c.\, c.} \right] \, - \, \delta Q_{\mu,i} \; .
\eea
But the integration above is, up to the volume factor, 
simply the average over the Fourier bin, hence we have:
\bea
\label{Eq:Qhat4}
\hat{Q}_{\mu,i} =
\frac{V_{\vec{k}_i}}{4 \, B_{\mu,i}^2 } 
\left\langle \tilde{f}_\mu \,
\tilde{f}^* + \; {\rm c.\, c.} \right\rangle_{\vec{k}_i}   \, - \, \delta Q_{\mu,i} \; .
\eea
Notice that, in this expression and others like it, the factor of 
$B_{\mu,i}^{-2}$ (which here plays the role of a normalization) 
is the {\em fiducial value} of the effective bias, 
whereas the weighted density contrasts $f_\mu$ must be computed 
directly from the data. However, since the weights of Eq. (\ref{Def:w}) are themselves
also computed using the fiducial values of $B_\mu$ and $P_m$, 
the weighted fields $f_\mu$ are a combination of both theory and data. 
The situation is not different from the usual case of Fourier analysis of 
cosmological surveys employing the FKP or the PVP estimators.
Evidently, these quadratic estimators are only truly
optimal if the parameters take their fiducial values.

Starting either from Eq. (\ref{Eq:Qhat4}), or more directly from Eq. (\ref{Def:Qhat}),
a long but straightforward calculation shows that the covariance of this
quadratic form in fact results in 
$Cov(\hat{Q}_{\mu,i},\hat{Q}_{\nu,j}) = F_{\mu,i \, ; \, \nu , j}$,
where the Fisher matrix was given in eq. (\ref{Eq:FBin}).

Finally we can construct the optimal quadratic estimators for the power spectra
of any tracer species, by plugging the quadratic form above into the appropriate
generalization of Eq. (\ref{Def:phat}). The Fisher matrix that is relevant in this case
was already given in Eq. (\ref{Eq:FBin}). We have, therefore, that the optimal quadratic
estimators, whose covariances are given by the inverse of the Fisher matrix, are given by:
\bea
\label{Eq:OptEst}
\hat{P}_{\mu,i} &=& \sum_\nu \sum_j \left[ F_{\mu,i \, ; \nu,j} \right]^{-1} \hat{Q}_{\nu,j}
\\ \nonumber
&=& \sum_\nu \left[ F_{\mu,i \, ; \nu,i} \right]^{-1} \hat{Q}_{\nu,i} \; ,
\eea
where the second line follows from the fact that the Fisher matrix is diagonal in
the phase space bins\footnote{This is only true if the Fourier-space bins
are sufficiently large, such that the spacing between them is larger than the
reciprocal of the typical size of the position-space bin, 
$\Delta k_i \gtrsim \pi/V_{\vec{x}_i}^{1/3}$ -- see, e.g., \citet{AbramoFisher}.} .

Now the origin of the normalizations of the weights, which
appear both in the FKP and the PVP formulas, becomes clear: up to the prefactor in 
Eq. (\ref{Eq:Qhat4}), those normalizations correspond to the inverse of the 
Fisher matrix in Eq. (\ref{Eq:OptEst}).


\subsection{Subtracting the bias of the estimators}

Although by definition 
$Cov(\hat{P}_{\mu,i},\hat{P}_{\nu,j}) = [F_{\mu,i \, ; \, \nu , j}]^{-1}$,
we still must ensure that the estimators are unbiased. According to Eq. (\ref{Eq:Dbias}), 
those biases are:
\bea
\label{Eq:Dmui}
\delta Q_{\mu,i} &=& \frac12 \sum_{\alpha\beta} 
\int d^3 x \, d^3 x' 
\frac{\partial C_{\alpha\beta} (\vec{x},\vec{x}{\,}') }{\partial P_{\mu,i}}
C^{-1}_{\beta\alpha} (\vec{x}{\,}',\vec{x}) 
\\ \nonumber
& & - \sum_\nu \sum_j F_{\mu,i \, ; \, \nu , j} P_{\nu,j} \; .
\eea
This expression can be easily worked out, and the result is:
\be
\label{Eq:dQmui}
\delta Q_{\mu,i} = \frac{1}{2 \, B_{\mu,i}^2} 
\int_{V_i} \frac{d^3 x \, d^3 k}{(2\pi)^3}
\frac{\bar{n}_\mu B_{\mu}^2}{1+{\cal{P}}}  
- \sum_\nu F_{\mu,i \, ; \, \nu , i} P_{\nu,i} \; .
\ee

It is also useful to compute the bias corrections for the power spectrum estimators,
$\hat{P}_{\mu,i}$.
For this calculation we will employ the approximation that averages over the bins
can be manipulated in such a way that
$\langle A \, B \rangle_i \simeq \langle A \rangle_i \, \langle B \rangle_i$.
This amounts to assuming that the bins are small compared with the coherence scale 
of the quantities of interest.

In order to go from $\hat{Q}_{\mu,i}$ to $\hat{P}_{\mu,i}$ we must first find 
the inverse of the Fisher matrix which was found in Eq. (\ref{Eq:FBin}). But that
is basically the inverse of the Fisher matrix for the $\log {\cal{P}}_\mu$ which
was found in Eq. (\ref{Def:Fcurved}). This is a particular case of the same type 
of matrix inversion which we used in the case of the pixel covariance, 
and the result is that:
\be
\label{Eq:IF}
F_{\mu\nu}^{-1} = P_{\mu} P_{\nu} \, {\cal{F}}_{\mu\nu}^{-1} \; ,
\ee
where:
\be
\label{Eq:IFC}
{\cal{F}}_{\mu\nu}^{-1} = \frac{4 (1+{\cal{P}})}{{\cal{P}} }
\left( \frac{\delta_{\mu\nu}}{ {\cal{P}}_\mu }
+ \frac{{\cal{P}}-1}{2{\cal{P}}}  \right) \; .
\ee
Using this expression, and the approximation that bin averages can 
be freely rearranged, we obtain that the estimators of the power spectra of
the tracers reduce to:
\be
\left\langle \hat{P}_\mu \right\rangle_i \rightarrow 
\left\langle \left(1+ \frac{1}{{\cal{P}}} \right) P_\mu \right\rangle_i - \delta P_{\mu,i} \; .
\ee
In fact, one can also show directly from Eq. (\ref{Eq:dQmui}) that the bias of the estimators are:
\be
\label{Eq:deltaP}
\delta {P}_{\mu,i} \equiv \sum_\nu [ F_{\mu,i \, ; \, \nu , i} ]^{-1} \delta Q_{\nu,i}
\, \rightarrow \,  \left\langle \frac{P_\mu}{{\cal{P}}} \right\rangle_i \; ,
\ee
which implies that $\langle \hat{P}_\mu \rangle_i \to P_{\mu,i}$, as it should.
In the case of a single type of tracer, the bias of the estimator reduces to the 
(Poissonian) shot noise, $1/\bar{n}$.


\subsection{The window functions}

The expectation values of the power spectra obtained through the multi-tracer
quadratic estimators are convolutions of the true 
power spectra with some window functions. These window functions can 
be obtained directly from the expectation value of the expression in
Eq. (\ref{Eq:Qhat}), by taking $\delta_\alpha \rightarrow B_\alpha \, \delta_m$ 
and neglecting the biases of the estimators:
\bea
\label{Eq:QWin}
\langle \hat{Q}_{\mu,i} \rangle &=&
\frac{1}{4 \, B_{\mu,i}^2 } 
\sum_{\sigma \alpha\beta}
 \int_{V_i} \frac{d^3 x \, d^3 k}{(2\pi)^3} \int d^3 x'
e^{i \vec{k} \cdot (\vec{x} - \vec{x}{ \, }')}
\\ 
\nonumber
& & \times 
w_{\mu\alpha} (\vec{x},\vec{k})  B_\alpha (\vec{x},\vec{k}) 
\,
w_{\sigma\beta} (\vec{x}{\,}',\vec{k})  B_\beta (\vec{x}{\,}',\vec{k}) 
\\ \nonumber 
& & \times \left\langle \delta_m (\vec{x}) \delta_m(\vec{x}{\,}')  \right\rangle + c.c. \; .
\eea
From the definition of the weight functions, Eq. (\ref{Def:w}), it is easy to derive that
$\sum_\sigma w_{\sigma\alpha} B_\alpha = {\bar{n}}_\alpha B_\alpha^2 / (1+{\cal{P}}) =
{\cal{P}}_\alpha / P_m (1+{\cal{P}})$, 
and expressing the matter 2-point correlation function in terms of the matter power
spectrum, we obtain:
\bea
\label{Eq:QWin2}
\langle \hat{Q}_{\mu,i} \rangle &=&
\frac{1}{4 \, B_{\mu,i}^2 } 
 \int_{V_i} \frac{d^3 x \, d^3 k}{(2\pi)^3} 
 \int \frac{d^3 x' \, d^3 k'}{(2\pi)^3} \, P_m(\vec{k}{\,}')
\\ 
\nonumber
& & \times 
e^{i (\vec{k}-\vec{k}{\,}') \cdot \vec{x}}
G_\mu ({\vec{x},\vec{k}}) \,
e^{-i (\vec{k}-\vec{k}{\,}') \cdot \vec{x}{\,}'}
G({\vec{x}{\,}',\vec{k}})
+ c.c. \; ,
\eea
where:
\be
\label{Def:Gx}
G_\mu ({\vec{x},\vec{k}}) = \frac{1}{P_m} \frac{{\cal{P}}_\mu}{1+{\cal{P}}} \; ,
\ee
and $G = \sum_\mu G_\mu={\cal{P}}/P_m(1+{\cal{P}})$.

Once again, one of the real-space integrals in Eq. (\ref{Eq:QWin2}) ought to be 
carried out only over the volume of
the spatial bin, $V_{\vec{x}_i}$, while the other should be in principle carried out over
the whole volume of the survey (e.g., all redshift slices). In practice, it may be
more conservative to treat each bin in position space as an entirely independent
survey, and in that case the two integrals over the real volume would be carried out
only on the volume of the bin $i$. In fact, it is only in this limit that 
the Fisher matrix of Eq. (\ref{Eq:FBin}), or that of Eq. (\ref{Eq:DefCurveF}), are
truly diagonal in the bins $i$ and $j$ \citep{AbramoFisher}, and therefore it is only
in this sense that the optimal estimators satisfy the constraint that 
$Cov(\hat{P}_{\mu,i},\hat{P}_{\nu,j}) \rightarrow [F_{\mu,i;\nu,j}]^{-1}$.

Hence, we define the window function:
\be
\label{Def:Window}
W_{\mu,i}^{(Q)} (\vec{k}_i,\vec{k}{\,}') = \frac{1}{4 B_{\mu,i}^2} \int_{V_i} \frac{d^3 k}{(2\pi)^3}
\tilde{G}_\mu (\vec{k},\vec{k}{\,}') \, \tilde{G}^* (\vec{k},\vec{k}{\,}')
\, + \, c.c.  \; ,
\ee
where the Fourier transform of the kernels of Eq. (\ref{Def:Gx}) are:
\bea
\label{Def:Gmu}
\tilde{G}_\mu (\vec{k},\vec{k}{\,}') 
& = &
 \int_{V_i} d^3 x 
\, e^{i (\vec{k}-\vec{k}{\,}') \cdot \vec{x}} \,
G_\mu(\vec{x},\vec{k}) \; ,
\\ 
\label{Def:G}
\tilde{G} (\vec{k},\vec{k}{\,}') & = &
\sum_\mu \tilde{G}_\mu (\vec{k},\vec{k}{\,}') \; .
\eea

Because the integral over $d^3 k$ is performed only over the Fourier bin 
$V_{\vec{k}_i}$, it is often an accurate approximation to take 
$\vec{k} \rightarrow \vec{k}_i$ in the argument of the kernels of Eqs. (\ref{Def:Gmu})-(\ref{Def:G}), and replace:
\bea
\label{Def:Gmui}
\tilde{G}_\mu (\vec{k},\vec{k}{\,}') 
& \rightarrow & \tilde{G}_{\mu,i} (\vec{k} - \vec{k}{\,}')
\\ \nonumber
& = &
 \int_{V_i} d^3 x 
\, e^{i (\vec{k}-\vec{k}{\,}') \cdot \vec{x}}
G_\mu (\vec{x},\vec{k}_i) \; ,
\\ 
\label{Def:Gi}
\tilde{G} (\vec{k},\vec{k}{\,}') 
& \rightarrow & \tilde{G}_i (\vec{k} - \vec{k}{\,}') =
\sum_\mu \tilde{G}_{\mu,i} (\vec{k} - \vec{k}{\,}') \; .
\eea
Notice that the Fourier transform of the kernels with respect to their spatial
dependence still remains, since we do not replace $\vec{k}$ by $\vec{k}_i$ in
the exponentials.

With these definitions, we can write the effective window function of the
quadratic form as:
\bea
\label{Eq:EffW}
W_{\mu,i}^{(Q)}
&=& 
\frac{1}{4 \, B_{\mu,i}^2}
\int_{V_i} \frac{d^3 k}{(2\pi)^3} 
\left[ \tilde{G}_\mu  \tilde{G}^*  + \, c.c. \, \right]
\\ \nonumber
& \simeq & 
\frac{V_{\vec{k}_i}}{4  \, B_{\mu,i}^2}
\left\langle
 \tilde{G}_{\mu, i} (\vec{k} - \vec{k}{\,}') \tilde{G}_i^* (\vec{k} - \vec{k}{\,}')
 \right\rangle_{V_{\vec{k}_i}} + \, c.c. \; .
\eea
Hence, in terms of the window function we have:
\bea
\label{Eq:QWin3}
\langle \hat{Q}_{\mu,i} \rangle \simeq
 \int \frac{d^3 k'}{(2\pi)^3} \, P_m(\vec{k}{\,}') \,
W_{\mu,i}^{(Q)} (\vec{k}_i,\vec{k}{\,'}) .
\eea
An interesting limiting case happens when we take all quantities to be 
constant inside the spatial bin, 
${\cal{P}}_\mu(\vec{x},\vec{k}) \rightarrow {\cal{P}}_{\mu,i} (\vec{k})$, 
and then take the continuum limit. In that case the kernels in Fourier space 
become Dirac-delta functions, 
$\tilde{G}_\mu \rightarrow G_{\mu,i} \, (2\pi)^3 \delta_D(\vec{k}-\vec{k}{\,}')$,
and the window function becomes:
\be
\label{Eq:WQlimit}
W_{\mu,i}^{(Q)} \rightarrow \frac{1}{4 \, B_{\mu,i}^2  P_{m,i}^2} \frac{{\cal{P}}_{\mu,i} {\cal{P}}_i}{(1+{\cal{P}}_i)^2} \times (2\pi)^3 \delta_D (\vec{k}_i-\vec{k}{\,}') \; .
\ee

The most relevant window functions are, of course, not $W^{(Q)}_\mu$, but
those which apply for the estimators of the power spectra. Since
$\langle \hat{P}_{\mu,i} \rangle 
= \sum_\nu [F_{\mu,i;\nu,i}]^{-1} \langle \hat{Q}_{\nu,i} \rangle$,
we obtain that:
\be
\label{Eq:PWin}
\langle \hat{P}_{\mu,i} \rangle 
=
 \int \frac{d^3 k'}{(2\pi)^3} \, P_m(\vec{k}{\,}') \, W_{\mu,i} 
\ee
where:
\be
\label{Eq:EffW}
W_{\mu,i}
= 
\sum_\nu [F_{\mu,i;\nu,i}]^{-1} 
\frac{1}{4 \, B_{\nu,i}^2 } 
\int_{V_i} \frac{d^3 k}{(2\pi)^3} 
\left[ \tilde{G}_\nu  \tilde{G}^*  + \, c.c. \, \right] \; .
\ee
Finally, in the same limit that was used to obtain Eq. (\ref{Eq:WQlimit}),
we can apply the identity:
\be
\sum_\nu F_{\mu\nu}^{-1} \times 
\frac{1}{B_{\mu}^2} \, \frac{1}{P_m} \frac{{\cal{P}}_\nu}{1+{\cal{P}}}
 =  2 B_{\mu}^2 \, P_m \frac{1+{\cal{P}}}{{\cal{P}}} \; ,
\ee
to show that 
$W_{\mu,i} \rightarrow  B_{\mu,i}^2 \times (2\pi)^3 \delta_D(\vec{k}_i - \vec{k}{\,}')$, 
as in fact it ought to be.
This completes the demonstration that the estimators derived in this Section
satisfy all the desired criteria for optimal, unbiased estimators, with the correct
continuum limits.


\subsection{Random maps and the integral constraints}
\label{Sec:Random}

Up to now we have introduced the optimal multi-tracer quadratic estimators without
mentioning the role of the random (``synthethic'') maps. They help subtract
the fluctuations that arise purely as a result of modulations in the mean number
density of the tracers, $\bar{n}_\mu(\vec{x})$, and are caused by, e.g., angular- or
redshift-dependent variations in the selection function of a survey 
\citep{1994ApJ...426...23F}.

For each tracer species with mean number density $\bar{n}_\mu(\vec{x})$
we define a random (white noise) map with a mean number density
with the same shape as that which is presumed for the data: 
$\bar{n}_\mu^r(\vec{x}) = \bar{n}_\mu(\vec{x}) /\alpha_\mu$,
where $\alpha_\mu$ are (small) constants.
The random datasets have no structure, in the sense that their pixel covariances
are just given by the shot noises of each sample:
\be
\nonumber
\langle \delta^r_\mu (\vec{x}) \delta^r_\nu (\vec{x}{\,}') \rangle 
= \frac{\delta_{\mu\nu}}{\bar{n}_\mu^r} \, \delta_D(\vec{x}-\vec{x}{\,}')
= \alpha_\mu \frac{\delta_{\mu\nu}}{\bar{n}_\mu} \, \delta_D(\vec{x}-\vec{x}{\,}') \; .
\ee

With the data and random sets we construct weighted density contrasts 
in a way similar to the definition of Eq. (\ref{Def:f}):
\bea
\label{Def:fr}
f_\mu (\vec{x},\vec{k}) 
&=& \sum_\nu w_{\mu\nu}(\vec{x},\vec{k}) \, 
\frac{ n_\nu(\vec{x}) - A_\nu \, n_\nu^r(\vec{x}) }{\bar{n}_\nu} \; ,
\\ \nonumber
&=& \sum_\nu w_{\mu\nu}(\vec{x},\vec{k}) \, 
\left[ \delta_\nu(\vec{x}) - \frac{A_\nu}{\alpha_\nu} \, \delta_\nu^r(\vec{x})  + 1 -  \frac{A_\nu}{\alpha_\nu} \right] \; ,
\eea
where the weights were given in Eq. (\ref{Def:w}).
The values of $A_\nu$ should be calibrated in such a way that the weighted fields
$f_\mu$ have zero mean over the volume of the sample, thus ensuring 
the so-called {\em integral constraints}, $\langle \hat{P}_\mu(k=0) \rangle \to 0$. 
It is easy to check that the condition $\int d^3 x \, f_\mu = 0 $ is satisfied by setting:
\be
\label{Eq:A}
A_\mu = \sum_\nu \, R_{\mu\nu}^{-1} \, D_\nu \; ,
\ee
where:
\bea
\label{Eq:Dnu}
D_\nu &=& \int 
d^3 x \sum_\sigma w_{\nu\sigma}(\vec{x},\vec{k}) \, \left[ 1 + \delta_\sigma(\vec{x}) \right] \; ,
\\ \label{Eq:Rmunu}
R_{\mu\nu} &=& \frac{1}{\alpha_\nu} \int 
d^3 x \; w_{\mu\nu} (\vec{x},\vec{k}) \, \left[ 1 + \delta_\nu^r (\vec{x}) \right]\; .
\eea
Since $D_\nu$ and $R_{\mu\nu}$ are functions of $\vec{k}$, in principle
the constants $A_\mu$ also depend on the wavenumber. 
In practice, we employ only a couple of 
putative values for $P_m$ in all the weights, hence we compute 
$A_\mu$ only for those values. 

Usually the mean density contrasts of the random catalogs
are very close to zero, which means that $A_\mu \to \alpha_\mu$ to a very 
good approximation. Indeed, taking $\delta_\nu^r \to 0$ in Eq. (\ref{Eq:Rmunu}) 
it follows that 
Eq. (\ref{Eq:A}) can be recast as:
\be
\label{ApproxA}
\frac{A_\mu}{\alpha_\mu} \approx  
1 + \sum_{\nu\sigma} \left[ \int d^3 x \; w_{\mu\nu} (x) \right]^{-1}
\int d^3 x' \; w_{\nu\sigma} (x')  \delta_\sigma (x') \; .
\ee
The fractional difference between $A_\mu$ and $\alpha_\mu$ is 
of the order of the average of the 
density contrast over the whole volume of the catalog.
This correction is negligible unless the galaxy catalogs are extremely sparse,
hence it is often safe to take $A_\mu \to \alpha_\mu$. One can also improve 
this approximation by taking smaller values of $\alpha_\mu$, which makes
Eq. (\ref{ApproxA}) more accurate. However, if there are reasons (e.g., computational)
to limit the size of the synthetic catalogs, such that $\alpha_\mu$ cannot be too small, 
then $A_\mu$ may deviate from $\alpha_\mu$.

Using Eq. (\ref{Def:fr}) instead of Eq. (\ref{Def:f}) 
in the estimators do not make much difference in our calculations,
except for the biases of the estimators, which inherit the factors of 
$A_\mu$ and $\alpha_\mu$. 
Starting from Eq. (\ref{Eq:dQmui}) we obtain:
\bea
\label{Eq:dQmuir}
\delta Q_{\mu,i} &=& \frac{1}{2 \, B_{\mu,i}^2} \int_{V_i} \frac{d^3 x \, d^3 k}{(2\pi)^3}
\frac{ \bar{n}_\mu B_{\mu}^2}{(1+{\cal{P}})^2}  
\\ \nonumber
& & \times \left\{ 1 + \sum_\nu \, \frac{A_\nu^2}{\alpha_\nu} 
\left[ \delta_{\mu\nu} (1+{\cal{P}}) - {\cal{P}}_\nu \right] \right\}
+ \Delta Q_{\mu,i}
\; ,
\eea
where the extra term, $\Delta Q_{\mu,i}$, arises when $A_\mu \neq \alpha_\mu$,
leading to the additional correction:
\bea
\label{DeltaQ}
\Delta Q_{\mu,i} 
& = &
 \frac{1}{2 \, B_{\mu,i}^2} \int d^3 x
 \sum_\nu \left(\frac{A_\nu}{\alpha_\nu} -1 \right) w_{\mu\nu} 
 \\ \nonumber
 & & \times
 \sum_{\gamma\sigma} \left(\frac{A_\sigma}{\alpha_\sigma} -1 \right) w_{\gamma\sigma}
 \; .
\eea
This expression can be simplified with the help of the definitions 
$\Gamma_\mu = (A_\mu/\alpha_\mu - 1) \bar{n}_\mu B_\mu$, and
$\Gamma = \sum_\mu \Gamma_\mu$, leading to:
\be
\label{DeltaQ2}
\Delta Q_{\mu,i} 
=  \frac{1}{2 \, B_{\mu,i}^2} \int d^3 x
\left[ \Gamma_\mu - \frac{\Gamma \, {\cal{P}}_\mu}{1+{\cal{P}}} \right] 
\frac{\Gamma}{1+{\cal{P}}}
\; .
\ee

As we discussed above, in most cases the tracers are sufficiently abundant to 
make $A_\mu \simeq \alpha_\mu$, so $\Gamma_\mu \to 0$, and the extra term of 
Eq. (\ref{DeltaQ2}) can be neglected. The biases of the estimators are then
given only by the first term of Eq. (\ref{Eq:dQmuir}) --- with the simplification
that $A^2_\mu/\alpha_\mu \to \alpha_\mu$.
If, in addition, we assume that the random maps are constructed such that 
the $\alpha_\mu$ are all identical, $\alpha_\mu \to \alpha$,
then the biases of the estimators become simply:
\be
\label{Eq:dQmuirsim}
\delta Q_{\mu,i} = \frac{1 + \alpha}{2 \, B_{\mu,i}^2} \int_{V_i} \frac{d^3 x \, d^3 k}{(2\pi)^3}
\frac{ \bar{n}_\mu B_{\mu}^2}{(1+{\cal{P}})^2}  
\; .
\ee


\section{Properties and relations of the multi-tracer estimator}
\label{Sec:Prop}

The results of the previous Section are closely related to other methods
for the Fourier analysis of cosmological surveys, but they also extend
their scope considerably. 

The simplest limit is when we take all tracers to be a single species.
In that case our formulas reduce to the ones by FKP. 
The weights of Eq. (\ref{Def:w}) reduce to $w=\bar{n} B/(1+{\cal{P}})$, which
are the FKP weights after we make the identification
${\cal{P}} = \sum_\mu {\cal{P}}_\mu = \bar{n} B^2 P_m$, where
$\bar{n} = \sum_\mu \bar{n}_\mu$ and  $B^2 = \bar{n}^{-1} \sum_\mu \bar{n}_\mu B_\mu^2$.
Furthermore, the multi-tracer Fisher matrix of Eq. (\ref{Def:Fcurved})
also reduces to the FKP Fisher matrix once we sum over all the clustering strengts 
--- i.e., when we combine all tracers into a single type. 
Changing variables in the Fisher matrix 
${\cal{F}}_{\mu\nu} = F [\log {\cal{P}}_\mu, \log {\cal{P}}_\nu]$,
from $\log {\cal{P}}_\mu$ to $\log {\cal{P}}$,
introduces a constant Jacobian, $J_\mu = \mathbb{1}_\mu$. This can be seen
by considering the inverse Jacobian, 
$J^{-1}_\mu = \partial \log {\cal{P}}/\partial \log {\cal{P}}_\mu = {\cal{P}}_\mu/{\cal{P}}$,
which satisfies $\sum_\mu J^{-1}_\mu J_\mu = \sum_\mu J^{-1}_\mu 
= \sum_\mu {\cal{P}}_\mu/{\cal{P}} = 1$.
Hence the multi-tracer Fisher matrix projected into the Fisher matrix for the total
clustering strength becomes:
\bea
F [\log {\cal{P}}] &=& \sum_{\mu\nu} J_\mu \, {\cal{F}}_{\mu\nu} \, J_\nu 
= \sum_{\mu\nu} {\cal{F}}_{\mu\nu} 
\\ \nonumber
&=& \frac12 \left( \frac{{\cal{P}}}{1+{\cal{P}}} \right)^2  \; ,
\eea
which is the FKP Fisher information density per unit of phase space volume.

We now discuss some of the main features of the multi-tracer technique, as well
as its relations to other methods in the literature.


\subsection{The PVP estimator}

Suppose we fix all parameters $B_{\mu,i}$, and try to estimate the matter
power spectrum $P_m(k)$ using data from all tracers. The optimal,
unbiased estimator in that case was derived by
\citet{Percival:2003pi} (PVP) --- see also \citet{SmithMarian15}. 
The method used by PVP to construct their 
estimator was the same as that used by FKP --- i.e.,
the weights which minimize the covariance $Cov(P_{m,i},P_{m,j})$ 
were obtained through a variational approach. 

Here, instead, we built the optimal estimators directly on the basis of the
pixel covariance, assuming Gaussianity of the data. 
We already showed that our estimators reduce to that of FKP
in the case of a single species of tracer. Now we show that the PVP estimator
is just one of many possible projections of the multi-tracer estimators.

If we fix the effective biases $B_{\mu}$ to their fiducial values (i.e.,
if the bias of each type of galaxy and the shape of the 
RSDs are set to their true values), then the 
remaining unknown is the matter power spectrum at the position- 
and Fourier-space bins, $P_{m,i}$. 
We may now ask what is the Fisher matrix for the matter power spectrum. This is
easily derived from Eq. (\ref{Eq:FBin}) through the change of variable:
\bea
\label{Eq:ProjFisher}
F(P_{m,i},P_{m,j}) &=& \sum_{\mu\nu} \sum_{kl} 
\frac{\partial P_{\mu,k}}{\partial P_{m,i}} \, 
F_{\mu,k;\nu,l} \,
\frac{ \partial P_{\mu,l}}{\partial P_{m,j}} 
\\ \nonumber
&=& \delta_{ij} \sum_{\mu\nu} 
B_{\mu,i}^2 \, B_{\nu,i}^2 \, 
F_{\mu,i;\nu,i} 
\\ \nonumber
&=& \frac{\delta_{ij}}{P_{m,i}^2}
\sum_{\mu\nu} {\cal{F}}_{\mu,i;\nu,i} 
\\ \nonumber
&=& \frac{\delta_{ij}}{P_{m,i}^2} \int_{V_i} \frac{d^3 x \, d^3 k}{(2\pi)^3} \, \frac12
\left( \frac{{\cal{P}}}{1+{\cal{P}}} \right)^2 \; ,
\eea
where we used that $\partial P_{\mu,k}/\partial P_{m,i} = B_{\mu,i}^2 \delta_{ki}$.
Hence, the Fisher matrix for the matter power spectrum is simply a 
projection of the multi-tracer Fisher matrix, where we {\em sum} the Fisher
information over all the tracers. Naturally, this result is also identical to what
was found in Eq. (\ref{Eq:FKP_P}) in the case of a single tracer --- i.e., in 
that case the PVP estimator reduces to the FKP estimator.

Now, if one fixes the effective biases and wishes to estimate the matter 
power spectrum alone,
then the generalization of Eqs. (\ref{Def:Qhat}) and (\ref{Def:Emu}) follow simply
by replacing the functional derivative
$\partial {\,} /\partial P_{\mu,i} \rightarrow \partial {\,} /\partial P_{m,i}$,
which is also equivalent to taking 
$\partial  {\,} /\partial P_{\mu,i} \rightarrow \sum_\mu B_{\mu,i}^2 \, 
\partial {\,} /\partial P_{\mu,i} $.
The resulting quadratic form is basically a projection of Eq. (\ref{Eq:Qhat4}):
\be
\label{Eq:PVP1}
\hat{Q}_{m,i}^{(PVP)} = \frac{V_{\vec{k}_i}}{2} 
\left\langle |\tilde{f}|^2 \right\rangle_{\vec{k}_i} \; ,
\ee
where the weighted field $f$ was defined in Eq. (\ref{ftot}).
Therefore, in the PVP estimator the density contrasts of all tracers
are combined into a single weighted density contrast, at each point in space. 
The cross-correlations are all averaged out, in such a way that
only the signal-to-noise of the matter power spectrum is optimized. 

The optimal estimator for the matter power spectrum is then simply
obtained by multiplying the quadratic form by the inverse of the Fisher matrix, i.e.:
\be
\label{Eq:PVP2}
\hat{P}_{m,i}^{(PVP)} = \frac{1}{N_i} 
\left\langle |\tilde{f}|^2 \right\rangle_{\vec{k}_i} \; ,
\ee
where the normalization is basically given by Eq. (\ref{Eq:ProjFisher}):
\be
\label{Eq:NormPVP}
N_i = \frac{1}{V_{\vec{k}_i} P_{m,i}^2} 
\int_{V_i} \frac{d^3 x \, d^3 k}{(2\pi)^3}
\left( \frac{{\cal{P}}}{1+{\cal{P}}} \right)^2 \; .
\ee
Noting that ${\cal{P}}/P_m = \sum_\mu \bar{n}_\mu B_\mu^2$, we see that
this estimator is precisely that of PVP.

One may ask also the converse question: what if we want to fix the matter 
power spectrum $P_m$, and estimate the effective biases 
$B_\mu$? In that case, it is a simple exercise to show that
this would lead right back to the optimal multi-tracer estimators, 
with the only difference that we would end up measuring 
$\hat{P}_{\mu,i}/P_{m,i}$. However, in reality we can only measure
the overall clustering of certain tracers of large-scale structure, which
means to estimate the combined product of the matter power spectrum 
and the (square of the) effective bias.
Any distinction between what belongs to the matter power spectrum, and
what belongs to the bias, RSDs, NGs, etc., can only be made after
some other type of prior knowledge is introduced -- e.g., by
constraining the normalization and shape of the spectrum from CMB 
observations, by modelling the RSDs, or by introducing priors on the bias 
from gravitational lensing.
Evidently, it would be an overuse of information to fix the power spectrum in
order to measure the bias, and then employ that bias in order to estimate
the power spectrum. Both are measured together in galaxy surveys, and
this fundamental degeneracy can only be broken by introducing additional 
data into the problem.


\subsection{The role of cross-correlations}

Although our estimators only compute the power spectra of the
individual tracers, $P_\mu = B_\mu^2 P_m$, it is clear from 
Eqs. (\ref{Eq:Qhat})-(\ref{Eq:Qhat2}) that the cross-correlations of the data,
$\langle \delta_\alpha \delta_\beta \rangle$ (with $\alpha \neq \beta$),
are also taken into account. 
In fact, the multi-tracer estimators express the optimal way to
combine both the auto- and the cross-correlations in the computation 
of the physical parameters $B_\mu$ and $P_m$.

Depending on the total signal-to-noise ratio (SNR), the power spectra 
of different tracers can have a positive or negative covariance. 
Since the SNR of a tracer is given by the amplitude of the power spectrum 
divided by shot noise, ${\cal{P}}_\mu = P_\mu/(\bar{n}_\mu)^{-1}$, the total 
SNR of a survey is expressed by the sum of the SNRs, ${\cal{P}}=\sum_\mu {\cal{P}}_\mu$.
Hence, when ${\cal{P}} \gg 1$ the total SNR is high, and conversely,
the total SNR is low if ${\cal{P}} \ll 1$.

When the total SNR is high, then from Eqs. (\ref{Eq:IF})-(\ref{Eq:IFC}) 
we immediately see that the covariance between the clusterings of  
different types of tracers (the off-diagonal terms) is positive, in fact 
$C_{\mu\nu} = Cov(P_\mu,P_\nu) \rightarrow 2 P_\mu P_\nu $. 
In relative terms, the covariance in that limit is constant for all tracers, 
$C_{\mu\nu}/(P_\mu P_\nu) \rightarrow 2 $. This is simply cosmic variance.

In the converse limit, of very low total SNR, the cross-covariance
becomes negative, 
$C_{\mu\nu} \rightarrow - 2 P_\mu P_\nu / {\cal{P}}^2$ ($\mu \neq \nu$).
In relative terms, the covariance in this limit is also independent of the
tracer species, as it happens in the high-SNR limit, but now 
$C_{\mu\nu}/(P_\mu P_\nu) \rightarrow -2/{\cal{P}}^2 $.


\subsection{Tracers with low SNR}

An obvious situation of interest arises when some tracer has low SNR. This can
happen if the tracer is sparse ($\bar{n}_\mu \lesssim 10^{-5}$), or has a very 
small bias ($b_\mu \ll 1$), making its clustering strength ${\cal{P}}_\mu$ 
very low in some bin or bandpower\footnote{Notice that the values of the 
power spectra of the tracers, $P_\mu=B_\mu^2 P_m$, should never actually vanish. 
If they do, in some sense (e.g., on extremely large or small scales), then 
${\cal{P}}\rightarrow 0$, making the entire Fisher matrix vanish for that bin --- as 
it should indeed happen in this case.}.
The danger would be that the inverse of the Fisher matrix (the covariance matrix),
which enters in the multi-tracer estimators through Eq. (\ref{Eq:OptEst}), 
could propagate this noise to the estimation of the spectra for the other tracers. 

However, this is not the case, as can be seen from the expression for the 
covariance matrix in Eqs. (\ref{Eq:IF})-(\ref{Eq:IFC}): because the reciprocals 
of the individual clustering strengths (i.e., the noises) only appear 
in the diagonal terms of the 
covariance matrix, if one of the tracers has a very high noise, this will only 
affect that same tracer. In particular, this means that our estimators are robust 
even when a galaxy survey includes tracers whose SNR are small.

This feature is very convenient if one would like to split a survey into several 
sub-surveys, by dividing galaxies, quasars and other objects into different categories 
according to type, luminosity, color, morphology, etc. -- all of which may be 
indicators of the bias of those tracers.
In doing that, even though the total SNR of the survey should remain 
approximately constant, the SNR of each individual tracer would decrease, 
leading us to wonder whether this could lead to a degradation of the 
information derived from that survey.
However, the fact that a tracer with low SNR only affects its own estimator 
means that this strategy can be safely used 
even when some tracers have very low number densities.


\subsection{Shot noise and the 1-halo term}

A fundamental assumption in our derivations has been that the covariance of the
counts of the tracers is given by Eq. (\ref{Eq:CovCounts}). However, this is often 
a simplification.

First, the statistics of counts in cells for galaxies in a redshift survey 
is only approximately Poissonian, so shot noise may be very different from 
the usual $1/\bar{n}_\mu$. 
Moreover, besides the 2-halo term which usually dominates on large scales,
there is an additional contribution to the power spectrum from the 1-halo 
term \citep{CooraySheth}. In the $k \to 0$ limit the 1-halo term is effectively
an additional contribution to shot noise.
In principle, any such corrections can be fixed simply by allowing
for a more general form of shot noise for each tracer which, in the
limit of negligible 1-halo term and Poisson statistics, 
reduces to $\delta_{\mu\nu}/\bar{n}_\mu$.

A closely related problem arises when different types of tracers occupy the same
dark matter halos. Eq. (\ref{Eq:CovCounts}) states that the covariance
between counts of different types of tracers do not have any shot noise. 
However, the Halo Model specifies that even
for galaxies of different types there is a non-vanishing 1-halo term, which
is degenerate with shot noise in the $k \to 0$ limit. 
Usually this is a small contribution, subdominant to the shot noise of the 
individual tracers, but it ultimately means that the noise cannot be
assumed to be diagonal in the tracers.

A third, and perhaps more serius problem, arises from that fact that
different tracers are often found to inhabit halos of very similar masses.
Most galaxies (as well as quasars)
are found in halos of masses in the range $10^{13} h^{-1} M_\odot 
\lesssim M_{h} \lesssim 10^{15} h^{-1} M_\odot $, with relatively small
differences between the distributions of each type of object within halos 
--- the so-called halo occupation distributions, or HODs
\citep{MartinezBook,CooraySheth}. 
In particular, this means that the biases of those tracers are not entirely 
independent. 

In other words, different tracers can be correlated by more than just 
the underlying dark matter field.
These correlations arise through the 1-halo terms of the power spectra,
which contribute to the covariances of the counts of those tracers, 
as well as through additional contributions to the bispectrum and trispectrum. 
But the trispectrum also defines the covariance of the power spectra through
$\langle P_\mu (\vec{k}) P_\nu (\vec{k}') \rangle
\sim T_{\mu\mu\nu\nu} (\vec{k},-\vec{k},\vec{k}',-\vec{k}')$, which means
that it is not possible to assume that the trispectrum is given only by the
connected pieces of the 4-point function --- i.e., it is not true anymore that
$\langle \delta_\mu \delta_\nu \delta_\alpha \delta_\beta \rangle =
C_{\mu\nu} C_{\alpha\beta} + C_{\mu\alpha} C_{\nu\beta} +
C_{\mu\beta} C_{\alpha\nu}$.

It is straightforward to incorporate the 1-halo term systematically into
the covariance in all our calculations (see Section \ref{S:1-halo}). However,
if there are significant correlations between the power spectra arising from 
the 1-, 2- and 3-halo terms of the trispectrum, then the counts cannot be 
assumed to be nearly Gaussian. In that case it would be erroneous to assume that 
the tracers are truly independent, and a key assumption of our method would be undermined. 
Nevertheless, \citet{SmithMarian15} were able to extend the PVP method (which does not
rely on a direct construction based on the pixel covariance, but on variational methods) to
incorporate these contributions from the Halo Model in formal expressions for the 
weights and for the Fisher matrix. However, recall that the PVP method, as well as its 
extension by \citet{SmithMarian15}, only tackle the estimation of the matter power 
spectrum, after assuming that the bias, RSDs, NGs, etc., are known and fixed.

\subsection{Degenerate tracers}

While tracers with different biases can possess correlations beyond 
those associated with the large-scale structure of the Universe, 
it is not necessarily true that two tracers that have similar 
biases must be highly correlated. 
Two types of galaxies may have different HODs, but their biases could coincide.
In those cases, if there is a significant contribution from the 1-halo term,
then it may still make sense to treat those species separately.
It is only when two tracers have the same HOD (or, equivalently, the 
same bias, 1-halo term, 2-halo term, 3-halo term, etc.), that they should 
be consolidated into a single species. 

However, suppose we do not know whether or not two types of galaxies have the same
HODs. If we use the multi-tracer approach and treat those two species as 
if they were different tracers, but they turn out to have the same HODs, would that 
initial assumption imply an overestimate of the information, or some distortion in the 
estimators?

The answer is no, and this follows from a very interesting property
of the multi-tracer Fisher matrix. As shown in \citet{AL13}, the Fisher
matrix can be diagonalized by changing variables, from the original power spectra
$P_\mu = B_\mu^2 P_m$ to a new set of parameters which correspond to
the total clustering strength and certain ratios between the power spectra --- 
the {\em relative clustering strengths}. In the case of two 
tracers with spectra $P_1$ and $P_2$, a choice of parameters which 
diagonalizes the Fisher matrix is $\log {\cal{P}} = {\cal{P}}_1 + {\cal{P}}_2$, 
and $\log {\cal{R}} = \log {\cal{P}}_1/{\cal{P}}_2$ (or, equivalently, 
$\log {\cal{P}}$ and $\log {\cal{P}}_2/{\cal{P}}_1 = - \log {\cal{R}}$).
The Fisher information per unit of phase space for this new set of parameters is:
\be
\label{DiagFish2}
{\cal{F}}[\log {\cal{P}}, \log {\cal{R}} ] = 
\left(
\begin{array}{cc}
\frac12  \frac{{\cal{P}}^2}{(1+{\cal{P}})^2} & 0 \\
0 & \frac14 \frac{{\cal{P}}^2 \, {\cal{R}}}{(1+{\cal{P}}) \, (1+{\cal{R}})^2}
\end{array} 
\right) \; .
\ee

For an arbitrary number $N$ of tracers, the change of variables that diagonalizes the 
Fisher matrix is identical to a change from Cartesian coordinates to spherical coordinates
in $N$ dimensions. Namely, if we regard the $N$ clustering strengths as 
${\cal{P}}_1 \to x_1^2$, ${\cal{P}}_2 \to x_2^2$, etc.,
then the variables that diagonalize the Fisher matrix are the radius, 
${\cal{P}} \to r^2 = \sum_\mu x_\mu^2$, together with the $(N-1)$ 
angles $\tan^2 \theta = (r^2-x_N^2)/x_N^2$, 
$\cot^2 \phi_1 = (r^2-x_N^2-x_{N-1}^2)/x_{N-1}^2$, 
$\cot^2 \phi_2 = (r^2-x_N^2-x_{N-1}^2-x_{N-2}^2)/x_{N-2}^2$, etc. 
Hence, the angle variables correspond to certain ratios between the tracers, 
(or {\em relative} clustering strengths),
for which the matter power spectrum (the radius) cancels out. 
In particular, this means that the 
relative clustering strengths are immune to some statistical limitations that affect 
the matter power --- namely, the relative clustering strengths can be measured to 
an accuracy which is not constrained by cosmic variance \cite{AL13}.

Coming back to our example of the two tracers, if we now stipulate that they 
are in fact a single species, then $P_1=P_2$, and ${\cal{R}} \to \bar{n}_1/\bar{n}_2$ 
is not a free parameter anymore, so $d \log{\cal{R}} \to 0$. 
This is equivalent to projecting the $2 \times 2$ Fisher matrix into 
a single component, thus eliminating the line and column corresponding to $\log{\cal{R}}$,
and leaving $\log {\cal{P}}$ as the sole free parameter. 
Indeed, since $d \log{\cal{R}} \to 0$ in this case, we cannot constrain physical parameters 
such as RSDs or NGs on the basis of a measurement of ${\cal{R}}$.

Since the Fisher matrix is diagonal, the Fisher information for $\log {\cal{P}}$ 
is unchanged after this projection (or marginalization).
In particular, the variance $\sigma^2(\log{\cal{P}})= \sigma^2({\cal{P}})/{\cal{P}}^2$ is 
untouched by a marginalization over ${\cal{R}}$, and it is still given by the inverse 
of the same Fisher matrix element in Eq. (\ref{DiagFish2}), so 
$\sigma^2({\cal{P}}) = 2 (1+{\cal{P}})^2$, which is nothing but the covariance 
(in units of phase space volume) for a single tracer species --- see Eq. (\ref{SigmaFKP}).

The argument above extends to any number of tracers: since the Fisher matrix is 
diagonal in the ``spherical coordinates'' (the total and relative clustering strengths), 
projecting some of the tracers out by combining them into new species does
nothing to the Fisher information of the total clustering strength, or to the
relative clustering strengths of the remaining species.
Therefore, in principle there is no difference between treating two identical
tracer species (with the same HODs) separately, or joining them into a single type of tracer.
Of course, one can always {\em destroy} information by treating two {\em different} 
tracer species as if they were just one, but there is no penalty for breaking 
a catalog into as many sub-catalogs as one wishes --- even if some of 
the tracers turn out to be completely degenerate.

The argument is a bit more involved if we work with the power
spectra as the parameters, but the conclusion is the same (see Appendix A).


\section{Testing the estimators}
\label{S:Applications}

In Sections 2 and 3 we derived the optimal multi-tracer estimators. 
We also obtained the covariance 
of the estimators --- which is simply the inverse of the multi-tracer Fisher matrix.
In this Section we apply that formalism to simple simulated galaxy maps. 
The implementation of the estimators is quite straightforward, and should be familiar 
to anyone who has used the FKP or the PVP methods. Although
we test the method in real space, the extension to redshift space is trivial: instead
of bins in $|\vec{k}|$, one should have bins both in $k$ and in $\mu_k^2$.

For the generation of the galaxy maps we chose a simple method
that is both efficient and computationally cheap enough that hundreds of realizations of 
a single fiducial matter power spectrum and galaxy model can be analyzed. 
We implemented the multi-tracer estimators in a cubic grid with constant, uniform 
mean number density (or selection function), for the case of two different 
species of tracers, with biases $b_1=1.0$ and $b_2=1.2$.
We checked that the estimators are as robust as the
FKP or PVP methods against variations in the survey geometry.

In order to test the performance of the estimators in situations of high or
low signal-to-noise, we consider three different cases, as shown in Table 1.
In each case we generate 1000 galaxy maps (each map consisting of two 
catalogs, one for each tracer), and estimate the spectra 
using the methods described in Sec. 3.

\begin{table}
\begin{center}
\begin{tabular}{|c|cccc|}
\hline
Case & $\bar{n}_1 \, (h^{3}$ Mpc$^{-3}$) & $b_1$ & $\bar{n}_2 \, (h^{3}$ Mpc$^{-3}$) & $b_2$ \\
\hline
A & $1. \, 10^{-2}$ & 1.0 &  $1. \, 10^{-2} $ & 1.2 \\
B & $1. \, 10^{-2}$ & 1.0 &  $1. \, 10^{-5} $ & 1.2 \\
C & $1. \, 10^{-5}$ & 1.0 &  $1. \, 10^{-5} $ & 1.2 \\
\hline
\end{tabular}
\label{Table:1}
\caption{The three cases we use to illustrate the application of the multi-tracer method. 
In all cases tracer 1 has bias $b_1=1.0$, and tracer 2 has bias $b_2=1.2$.
In case A the two tracers have high number densities, so the signal-to-noise is high.
In case B tracer 1 is dense, but tracer 2 is sparse.
In case C both tracers are sparse, so the signal-to-noise is low.}
\end{center}
\end{table}

\subsection{Lognormal maps}

Our mocks follow the same procedure used in, e.g., PVP. A detailed 
description of the generation of lognormal maps can be found in \citet{ColesJones91}. 
The basic idea is that a Gaussian density contrast $\delta^{(G)}(\vec{x})$ is not bounded
from below, which implies that negative values for the density are possible in any 
finite-volume realization of such a Gaussian field. Lognormal fields, on the other 
hand, are positive-definite, so we map the Gaussian field into a lognormal field.

A lognormal field obeys the condition $\delta^{(L)}(\vec{x}) \geq -1$ and approximately 
describes the non-linear density field at low redshifts. We can obtain a 
lognormal density field in terms of a Gaussian density field through 
the definition $1+\delta^{(L)} (\vec{x})= \exp[ \delta^{(G)}(\vec{x})- \, \sigma_G^{2}/2]$, 
where $\sigma_G^{2}$ is the variance of the Gaussian field inside a cell. 
The Gaussian correlation function is related to the physical (assumed lognormal) 
correlation function
by $\xi^{(G)}(x)=\ln[1+\xi^{(ph)}(x)]$. Given a fiducial cosmology, we obtain the 
$z=0$ matter power spectrum $P_m(k)$ from the Boltzmann code \texttt{CAMB}
\footnote{http://CAMB.info} \citep{CAMB}, 
and inverse-Fourier transform it to get the physical correlation function $\xi^{(ph)}(x)$. 
We then convert the physical (assumed lognormal) correlation function to the 
correlation function of the corresponding Gaussian field, and Fourier-transform 
that correlation function into a power spectrum for the Gaussian field. This is 
the power spectrum which is employed 
to generate the Gaussian random modes for the density contrast.

The next step is the generation of biased lognormal maps for each galaxy type. 
We {\em define} the lognormal maps as
$1+\delta^{(L)}_\mu (\vec{x})= \exp[ b_\mu \, \delta^{(G)}(\vec{x})- \, b_\mu^2 
\sigma_G^{2}/2]$ \footnote{Notice that, for a lognormal map with bias $b$,
the correlation function used in the generation of the Gaussian random modes
should be defined as $\xi^{(G)}(x)= b^{-2} \, \ln[1+ b^2 \, \xi^{(ph)}(x)]$. 
Therefore, strictly speaking, this prescription only is self-consistend when 
there is a single type of galaxy, with one bias.
However, using the same correlation function for tracers of different biases
introduces only a small spectral distortion on small scales, which we corrected for 
in our simulations.} .
Finally, we create the galaxy maps as independent Poisson realizations over the 
lognormal fields. Each tracer has its own spatial 
number density $\bar{n}_\mu (\vec{x})$ and 
bias $b_\mu$, so that the maps for each tracer are given by integer numbers 
for each cell of volume $dV$ in our cube through a Poisson sampling,
$N_\mu (\vec{x}) \leftarrow \mathbb{P}\{\bar{n}_\mu (\vec{x}) [1+\delta^{(L)}_\mu(\vec{x})] dV\}$, 
where $\mathbb{P}\{\lambda\}$ is a Poisson distribution with mean $\lambda$.

In the three cases detailed above we 
considered cubic $256^3$ grids with a fiducial cosmology characterized by a flat 
$\Lambda$CDM model with $\Omega_{b}h^{2}=0.0226$, 
$\Omega_{CDM}h^{2}=0.112$ and $h=0.72$. Each cube
has a physical (comoving) volume of (1280$\,h^{-1}$Mpc)$^3$. It is important to note that 
lognormal maps created this way do not show the usual effect of suppression 
in power at small scales when a smoothing algorithm is applied to convert from 
a continuous distribution to a discrete grid, such as Nearest Grid Point (NGP). 
In any case, the formalism is general enough to accommodate this necessity. 
Furthermore, since the grid used is cubic, it is unnecessary to deconvolve the 
estimated spectra from the window function. Even though any discretization scheme
could be used, the square grid is required in order to employ
an implementation in terms of a fast Fourier Transform (FFt), which is, as a matter of fact, 
the only practical way to perform a Fourier analysis of large data sets.

\subsection{The data analysis algorithm}
\label{SubSec:Algo}

With the galaxy maps $n_{\mu}(\vec{x})$ as input, along with 
an initial guess for the biases $b_{\mu}$, we can start to deploy the machinery 
developed in Secs. 2 and 3. A previous step, in case we had not explicitly 
generated maps with constant, uniform number densities, would be to 
estimate $\bar{n}_\mu(\vec{x})$.

We start by constructing random maps, $n_\mu^r(\vec{x})$, for each tracer
as a Poisson process, in each cell of the grid, with the same shape for the 
mean number density as the data (i.e., the real maps), but with a larger number of particles,
$\bar{n}_\mu^r = \bar{n}_\mu /\alpha_\mu$, where $\alpha_\mu$ are small constants. 
We then construct the density contrasts according to Eq. (\ref{Def:fr}):
$\delta_{\mu}(\vec{x})= (n_{\mu}^{d} - A_\mu \, n_{\mu}^r)/\bar{n}_{\mu} $
 --- where recall that $A_\mu$ are constants found
according to the discussion in Sec. \ref{Sec:Random}.

\begin{figure}
\resizebox{9.0cm}{!}{\includegraphics{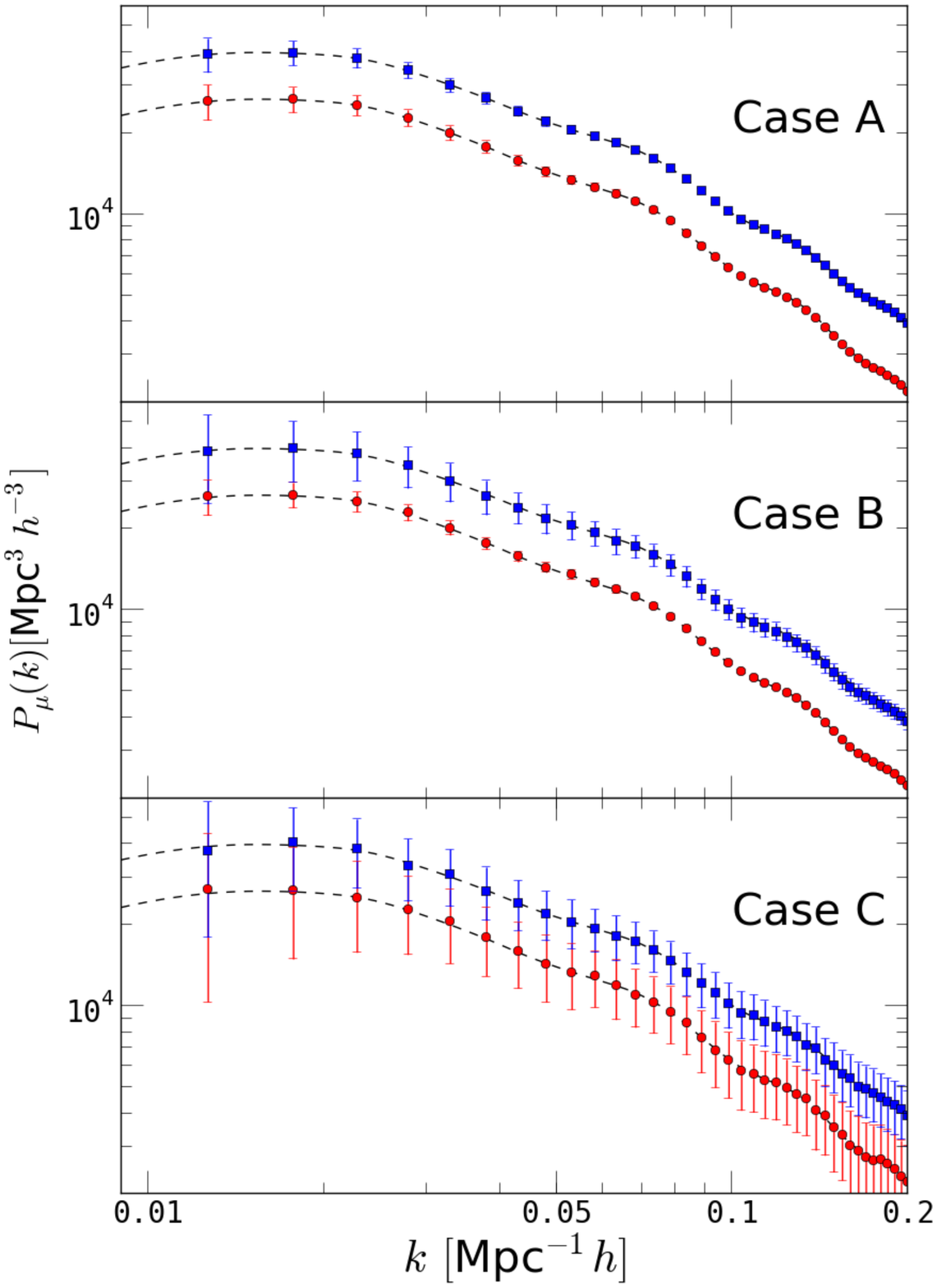}}
\caption{Estimated auto-spectra {\it v.} real auto-spectra. 
Filled (red) circles correspond to the power spectrum of the 
tracer 1, with $b_1={1.0}$, and filled (blue) squares 
correspond to the estimated power spectrum of the tracer 2, 
with bias $b_2=1.2$.
The symbols and error bars correspond to the mean and to the variance,
respectively, of 1000 realizations.
The dased lines are the input (theoretical) spectra of the tracers, given 
their biases and our fiducial cosmology. 
The upper, middle and lower panels correspond to cases A, B and C, 
respectively (see Table 1). 
The error bars are the 
theoretical ones --- i.e., the inverse of the Fisher matrix, Eq. (\ref{Eq:FBin}).}
\label{Fig:PcaseABC}
\end{figure}

With an initial guess for the biases
and for the amplitude of the power spectrum, we can 
construct ${\cal{P}}_\mu$ and ${\cal{P}}=\sum_\mu {\cal{P}}_\mu$, 
plug them into the weights (\ref{Def:w}), 
and calculate the weighted density constrasts of Eq. (\ref{Def:fr}). 
We then perform an FFt over $f(\vec{x})$ and 
$f_\mu(\vec{x})$, in order to obtain the integrand of Eq. (\ref{Eq:Qhat3}).
Taking proper care of the volume factors (in real and in Fourier space), 
this step should be analogous to the average over modes in Eq. (2.4.5) of FKP.

The next step is to subtract the biases of the estimators 
--- the $\delta Q_{\mu,i}$ in Eq. (\ref{Eq:Qhat3}) or, equivalently, Eq. (\ref{Eq:dQmuirsim}). 
Assuming that averages over bins
are such that $\langle A \, B \rangle_i \approx \langle A \rangle_i \, \langle B \rangle_i$,
and taking a single value for all the $\alpha_\mu \to \alpha$,
Eq. (\ref{Eq:dQmui}) can be rearranged to yield:
\be
\label{Eq:SdQmui}
\delta Q_{\mu,i}  =  \frac{1 + \alpha}{2} 
\int_{V_i} \frac{d^3 x \, d^3 k}{(2\pi)^3}
\frac{\bar{n}_\mu}{(1+{\cal{P}})^2} \; .
\ee
With our choice of $\alpha = 10^{-6}$, we find that $A_\mu \to \alpha$ to an
excellent approximation, which means that the biases of the estimators are
given only by Eq. (\ref{Eq:SdQmui}) --- see Sec. \ref{Sec:Random}.
Finally, the estimated power spectra are computed with the help of
Eq. (\ref{Eq:OptEst}).

We present our results for the estimated spectra of two types of tracers
in three cases, A, B and C --- see Table 1 and Fig. \ref{Fig:PcaseABC}. 
Case A represents a low-redshift survey which is highly complete, so
both tracers are dense.
Case B represents a low- or intermediate-redshift survey, with
one dense species of tracer (type 1 --- say, red galaxies) and one sparse 
species of tracer (type 2 --- say, quasars). Case C represents
a high-redshift survey, with two sparse types of tracers.

Our estimates were evaluated in evenly separated bandpowers with 
$\Delta k = 0.005 \, h$ Mpc$^{-1}$. We show the estimated spectra
in Fig. \ref{Fig:PcaseABC}, only up to $k = 0.2 \, h$ Mpc$^{-1}$ --- slightly into the 
nonlinear regime but still below the Nyquist frequency, such that our results are 
not affected by discretization effects.
When estimating the spectra we adopted a 
commonly used simplification, which is to fix the value of the matter power spectrum 
that is used in the weights, Eq. (\ref{Def:w}) 
--- in our case, we found that fixing $P_m \to 10^4 \, h^{-3}$ Mpc$^3$ in the
weights was a suitable choice.
Our results did not change significantly over the dynamical range of interest
when that value was multiplied by 2 or by 1/2.

\subsection{Empirical {\it v.} theoretical covariances}

We now check whether the theoretical covariance matrix (the inverse of the
multi-tracer Fisher matrix) is a good approximation to the true (i.e., empirical)
covariance matrix. If the theory is accurate, then the method is validated; 
if it is not, then the multi-tracer estimators are sub-optimal.

The empirical result was obtained from 1000 realizations. This was
compared with the theoretical covariance --- 
i.e., the inverse of the binned Fisher matrix of  Eq. (\ref{Eq:FBin}):
\be
\label{RelCov}
Cov(P_{\mu,i},P_{\nu,j}) = \delta_{ij}
\left[ \frac{1}{P_{\mu,i} \, P_{\nu,i}} 
\int_{V_i} \frac{d^3 x \, d^3 k}{(2\pi)^3} {\cal{F}}_{\mu\nu} \right]^{-1}
\; ,
\ee
where ${\cal{F}}_{\mu\nu}$ was defined in Eq. (\ref{Def:Fcurved}).

\begin{figure}
\begin{center}
\resizebox{8.5cm}{!}{\includegraphics{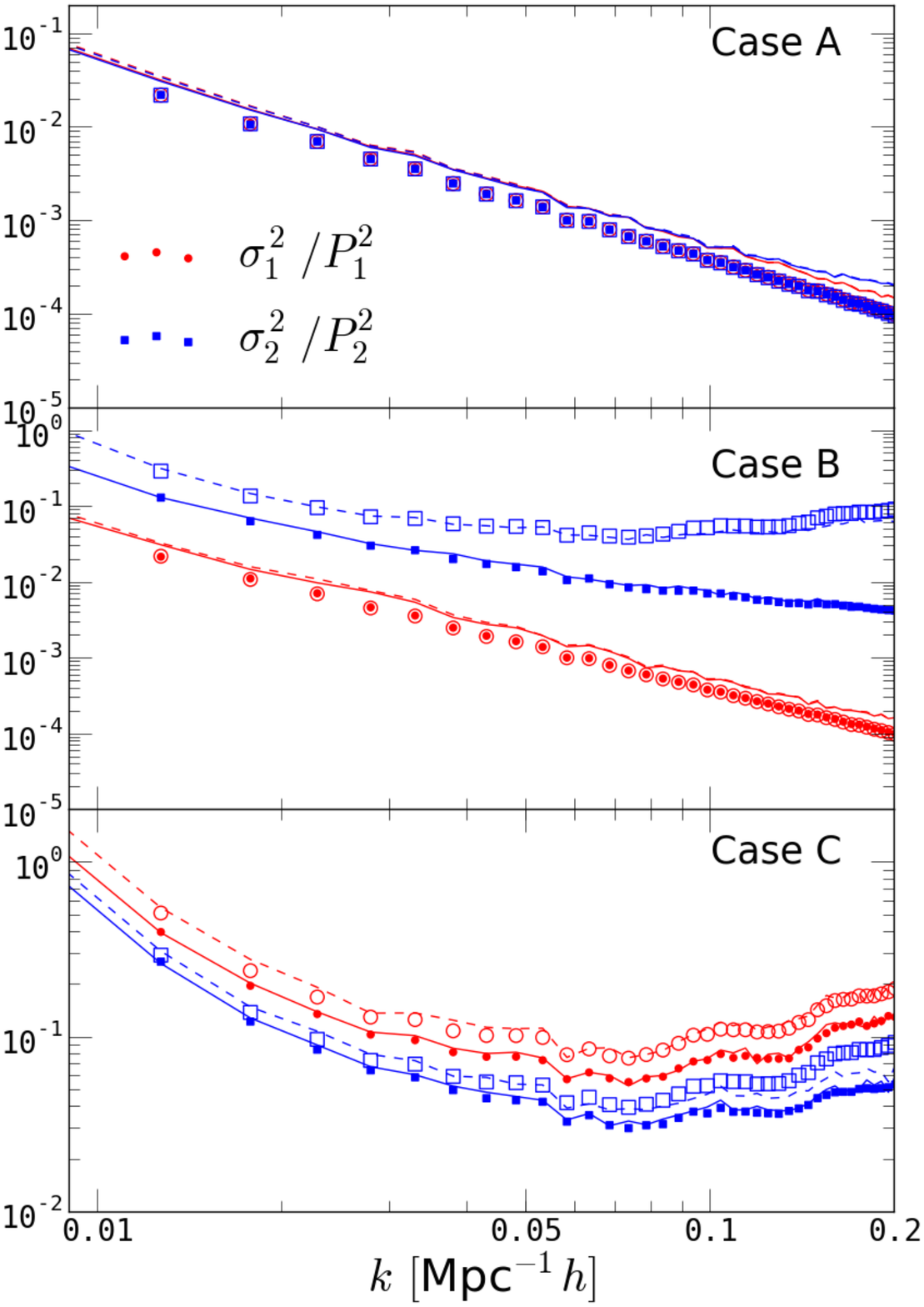}}
\caption{
Theoretical {\it v.} empirical relative covariances 
of the auto-spectra, $Cov(P_{\mu,i},P_{\nu,i})/P_{\mu,i} P_{\nu,i}$.
The upper, middle and lower panels correspond to cases A, B and C, respectively 
(see Table 1).
Red circles and blue squares correspond 
to the theoretical covariances of the tracers 1 and 2, respectively. 
The lines of the same colors are the standard deviation of our 1000 lognormal mocks. 
Solid symbols and lines correspond to multi-tracer estimates, while
open symbols and dashed lines correspond to FKP estimates.
In case A (upper panel), since the two tracers have high signal-to-noise 
(both ${\cal{P}}_1 \gg 1$ and ${\cal{P}}_2 \gg 1$ in
this range of scales), both the multi-tracer and the FKP formulas for the auto-covariances 
reduce to $Cov(P_{\mu,i},P_{\nu,i})/P_{\mu,i} P_{\nu,i} \simeq 2/V_{x,i} V_{k,i} \sim k^{-2}$
 [see Eqs. (\ref{Eq:IF})-(\ref{Eq:IFC})]. Hence, in this case most symbols and lines overlap.
In most cases, the empirical covariances are slightly higher than the theoretical ones
--- as expected. In case B (middle panel), 
the covariance of spectrum of the sparse tracer species is 
significantly higher in the FKP method: in this case, the multi-tracer method reduces the
uncertainty in the spectrum by a large factor.}
\label{Fig:cov_caseABC.png}
\end{center}
\end{figure}

\begin{figure}
\resizebox{8.85cm}{!}{\includegraphics{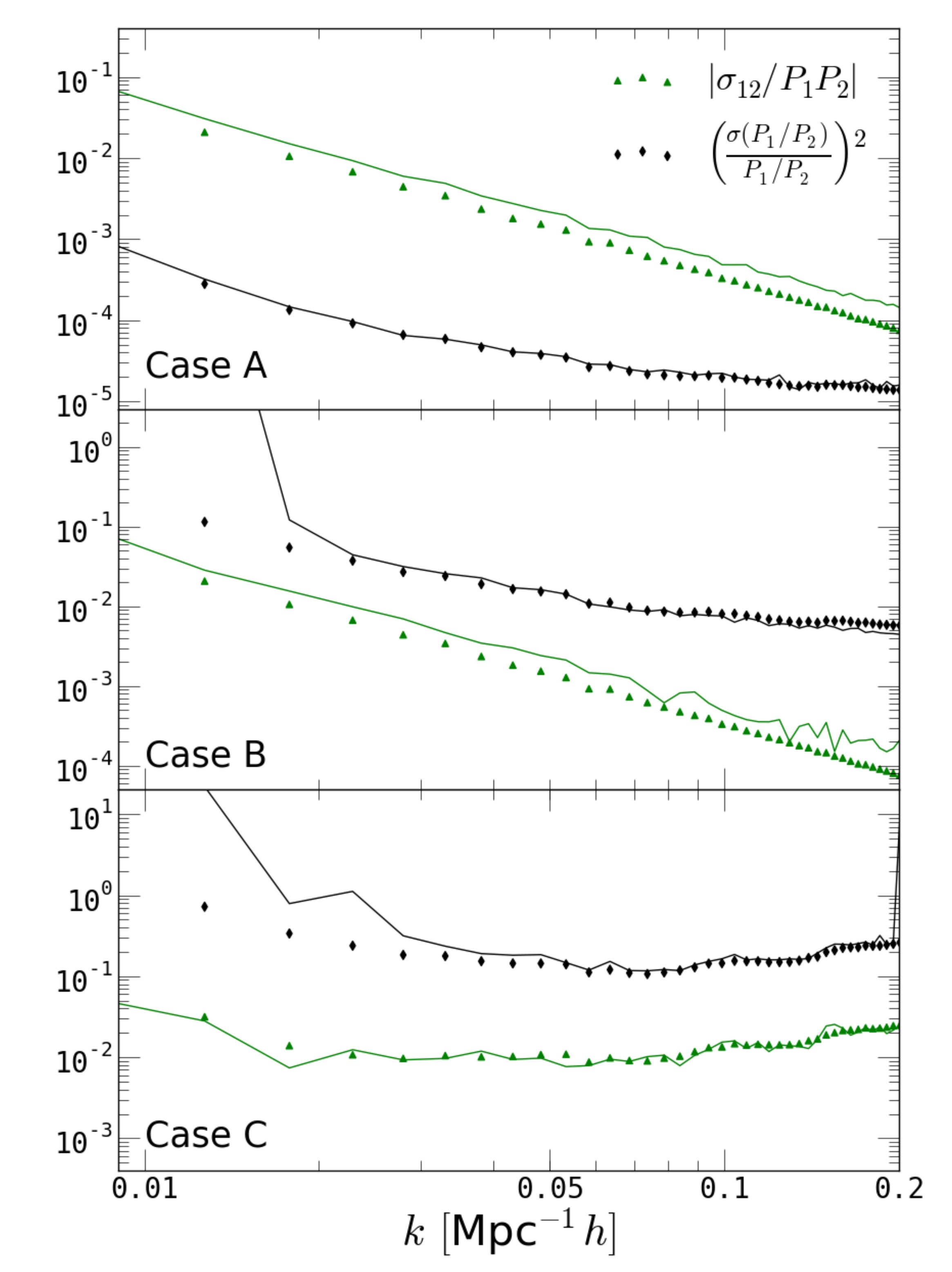}}
\caption{Theoretical {\it v.} empirical covariances of the cross-spectra
and of the ratios between the spectra.
The ratios were defined as $P_1/P_2$ (the relative covariance is identical for $P_2/P_1$).
The upper, middle and lower panels correspond to cases A, B and C, respectively 
(see Table 1).
Diamonds (black) correspond to the theoretical relative covariances of the
cross-spectra, while triangles (green) correspond to the theoretical covariance
for the ratios between the spectra (see text).
The solid lines correspond to the empirical covariances, using the multi-tracer estimators
(we do not show the results using the FKP estimator in these plots because it 
performs significantly worse compared with the multi-tracer estimators, and in any case
the FKP method does not predict these covariances).
Notice that in case C (lower panel) the covariance of the cross-correlations 
is negative, since ${\cal{P}} < 1$ --- see Eqs.(\ref{Eq:IF})-(\ref{Eq:IFC}).
Notice also that in case A the ratio between the spectra has a much lower uncertainty
than the cross-correlation (for an explanation, see the text).}
\label{Fig:cross_ABC}
\end{figure}

\begin{figure}
\resizebox{9.0cm}{!}{\includegraphics{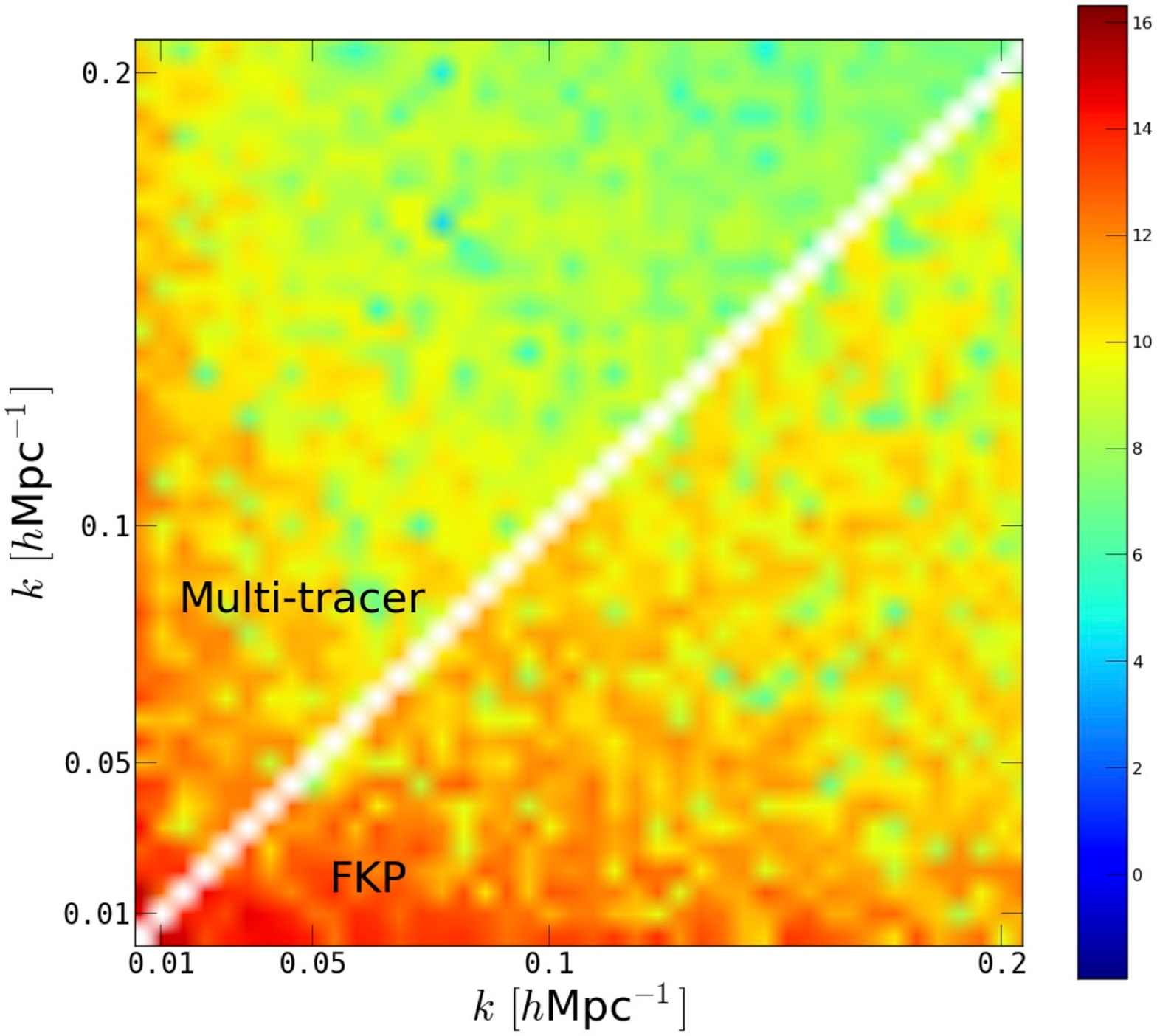}}
\resizebox{9.0cm}{!}{\includegraphics{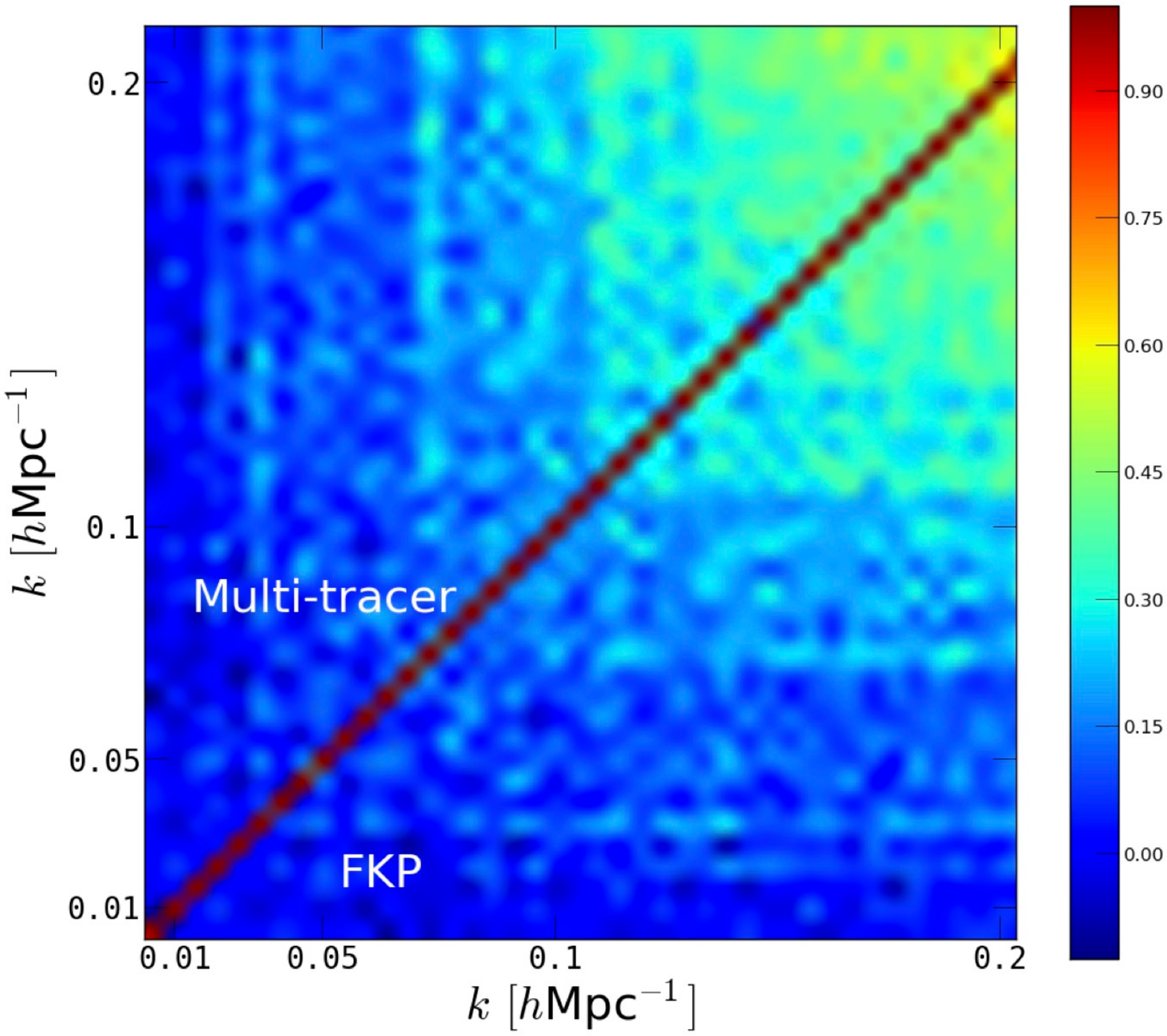}}
\caption{{\bf Upper panel}: covariance matrix for tracer 2 in case B. The 
upper triangle is the result using the 
multi-tracer estimator, and the lower triangle results from using the FKP estimator. 
The multi-tracer technique performs significantly better on all scales.
{\bf Lower panel}: correlation matrix for tracer 2 in case A. Both the multi-tracer and the FKP
estimators perform similarly regarding the correlations between Fourier bins. 
We checked that the estimators result in similar correlation matrices for both 
tracers, in the three different cases we analyzed.}
\label{Fig:corr}
\end{figure}

In Fig. \ref{Fig:cov_caseABC.png} we present the comparison 
between the theoretical and empirical covariances for the auto-spectra of the two species, 
obtained respectively from Eq. (\ref{RelCov}) and from taking the standard deviation of 
1000 lognormal realizations. We find that our theoretical expression properly reproduces 
the behavior of the statistical fluctuations in all cases, 
matching more closely the variances when compared with the FKP method.
The theoretical variances sometime underestimate slightly the empirical variance,
which is consistent with the notion that the inverse of the Fisher matrix is an underestimate 
of the true covariance. This is in line with what is usually found in implementations of the 
FKP method. In cases B and C the multi-tracer estimator performs significantly better 
than the FKP estimator on all scales.

In Fig. \ref{Fig:cross_ABC} we compare the theoretical and empirical variances 
for the cross-spectra of the two tracers (green triangles), and for
the ratios of the two spectra, $P_1/P_2$ (black diamonds).
Since the FKP method cannot predict theoretical covariances in these two cases,
we only show the multi-tracer theoretical variances.
The theoretical variance for the ratio $P_1/P_2$ follows from the 
multi-tracer Fisher information matrix, Eq. (\ref{Def:Fcurved}), which can be diagonalized 
by a change of variables \citep{AL13}, where the new parameters (the ``eigenvectors'' of
the Fisher matrix) are not the individual clustering strengths
${\cal{P}}_\mu$, but the total clustering strength, ${\cal{P}}$, and certain ratios between the 
clustering strengths. In particular, a diagonal Fisher matrix means that the degrees of freedom 
are independent --- there are no cross-covariances.
For two types of tracers, the variables which
diagonalize the $2\times2$ Fisher matrix are ${\cal{P}}={\cal{P}}_1+{\cal{P}}_2$,
and ${\cal{P}}_1/{\cal{P}}_2$ (or, equivalently, ${\cal{P}}$ and ${\cal{P}}_2/{\cal{P}}_1$). 
As shown in \citet{AL13}, the Fisher matrix per unit of phase space volume for
$\log(P_1/P_2)$ is $F_{\rm ratio} = {\cal{P}}_1 \, {\cal{P}}_2/4(1+ {\cal{P}}_1 + {\cal{P}}_2)$, 
from which follows that the relative covariance of that ratio is  
$ [\int d^3 x \, d^3 k/(2\pi)^3 F_{\rm ratio}]^{-1}$. This figure demonstrates the power of the
multi-tracer technique to measure $P_1/P_2 = B_1^2(z,k,\mu_k)/B_2^2(z,k,\mu_k)$,
something that can be used to place stronger constraints not only the biases of the 
two species, but also on RSDs, NGs, etc.

The upper panel of Fig. \ref{Fig:corr} shows the covariance matrix for tracer 
2 ($b_{2}=1.2$) in case B --- i.e., $Cov_{22}^{(B)}(k_i,k_j)$.
We exploited the symmetry of the covariance matrix under $k_i \leftrightarrow k_j$ 
in order to compare the multi-tracer and FKP estimators directly. 
In the lower panel of this figure we show the correlation matrix, defined as
Corr$_{ij}=Cov_{ij}/\sqrt{Cov_{i i} \, Cov_{j j}}$. We find that both the multi-tracer 
and the FKP estimators yield roughly similar correlation matrices, with weakly 
correlated bins up to scales $k \lesssim 0.1 \,h$ Mpc$^ {-1}$.

The upper panel of Fig. \ref{Fig:corr}, together with the middle panels of 
Figs. \ref{Fig:cov_caseABC.png} and  \ref{Fig:cross_ABC}, 
shows that in case B the multi-tracer estimator performs significantly better than the 
FKP estimator at all scales, with uncertainties up to one order of magnitude
smaller for the spectrum of the sparse tracer. 
The multi-tracer technique is also clearly superior in estimating the auto-spectra
in case C, when both tracers are sparse --- see the lower panel of Fig. \ref{Fig:cov_caseABC.png}.


\section{Including the 1-halo term}
\label{S:1-halo}

The fundamental object in this paper, which was used to derive the Fisher information 
matrix, as well as the optimal weights, is the pixel covariance. In the limit where bias and
RSDs depend weakly on $\vec{k}$, the covariance can be approximated by 
Eq. (\ref{Eq:CovLim}). However, this is not a complete description: in addition to the 
``signal'', $P_\alpha = B_\alpha^2 P_m$, and the shot noise, 
$\delta_{\alpha\beta}/\bar{n}_\alpha$,
there is another source of correlations between the density contrasts of different species
of tracers at different points in space: the 1-halo term of the power spectrum. 
According to the Halo Model
\citep{CooraySheth}, dark matter halos are the genuine tracers of the underlying 
matter density, while galaxies only trace the halos. In particular, this means that many
galaxies may be hosted by the same halo, in which case they would be tracing the 
same features of the underlying fluctuations of the matter density.

This additional covariance between galaxy counts is expressed by the 1-halo term:
\be
\label{P1h}
P^{1h}_{\alpha\beta} (k) = \frac{1}{\bar{n}_\alpha \, \bar{n}_\beta}
\int d \ln M \, \frac{d \, \bar{n}_h}{d \ln M}  \, u^2(k|M) \, \langle N_\alpha N_\beta \rangle_M \, 
  \; ,
\ee
where $d \, \bar{n}_h/d \ln M$ is the mass function for halos of mass $M$, 
$N_\alpha$ is the number of galaxies of type $\alpha$, and
$u(k|M)$ is the Fourier transform of the halo profile \citep{CooraySheth}.
The expectation value is over the probability distribution function for the
numbers of galaxies (the HOD) at a given halo mass. 
For the species of tracers which are typically used in cosmological surveys 
the 1-halo term is only relevant on small scales ($k \gtrsim 1 \, h$/Mpc) 
--- although, since $u(k\to0) =1$, it still contributes a constant factor on large scales.

Inclusion of the 1-halo term would lead the approximated pixel 
covariance of Eq. (\ref{Eq:CovLim}) to assume the expression:
\be
\label{PixCovP1h}
C_{\alpha\beta} (\vec{x},\vec{x}{\,}') \rightarrow \delta_D (\vec{x},\vec{x}{\,}') 
\times \left[ \frac{\delta_{\alpha\beta}}{\bar{n}_\alpha} + P^{2h}_{\alpha\beta}
+ P^{1h}_{\alpha\beta} \right] \; ,
\ee
where we write the 2-halo term $P^{2h}_{\alpha\beta} = B_\alpha B_\beta P_m$.
In principle, if we are only interested on the properties of the clustering on large scales, 
this term can be included systematically, in every step of the calculations
--- see also \citet{Hamaus2010}, in a similar context. 
These are straightforward computations, but for a general
form of $P^{1h}_{\alpha\beta}$ there is no closed-form expression for the
inverse of the pixel covariance matrix, which means that we cannot give
explicit formulas for the Fisher matrix, the weights, the window functions, etc.


\subsection{Fisher matrix of the 2-halo term for separable 1-halo terms}

In some cases the populations of tracers are such that the 1-halo term
is approximately separable, i.e., it can be expressed as a direct product of two terms, 
$P^{1h}_{\alpha\beta} \sim H_\alpha H_\beta$ --- just as happens with the 2-halo term.
We have checked that, for a class of HODs that is commonly used to 
describe red and blue galaxies \citep{Zheng2005}, all entries of the correlation matrix 
$P^{1h}_{\alpha\beta}/\sqrt{P^{1h}_{\alpha\alpha} P^{1h}_{\beta\beta} }$ are very close
to unity, which justifies this approximation. However, we only verified this feature 
of the 1-halo term while ignoring the distinction between central and satellite 
galaxies, since it is not clear how to generalize $\langle N_\alpha N_\beta \rangle$
in that case. It would be interesting to find out whether this property holds for more
realistic HODs.

If the 1-halo term is separable, it turns out that we {\em can} invert the covariance matrix.
This result follows from the exquisite properties of matrices that can be written as
$M_{\alpha\beta} = \delta_{\alpha\beta} + v_\alpha v_\beta + u_\alpha u_\beta$.
This type of matrix appeared already in Section 2, where we showed that the
inverse of $M_{v,\alpha\beta} = \delta_{\alpha\beta} + v_\alpha v_\beta$ is given
by $M_{v,\alpha\beta}^{-1} = \delta_{\alpha\beta} - v_\alpha v_\beta /(1+v^2)$, where
$v^2 = \sum_\mu v_\mu^2$.

As shown in Appendix B, the inverse of the matrix 
$M_{\alpha\beta} = \delta_{\alpha\beta} + v_\alpha v_\beta + u_\alpha u_\beta$ is:
\be
\label{Ap:InvM}
M^{-1}_{\alpha\beta} =  \sum_{\mu\nu} M_{v,\alpha\mu}^{-1/2} 
\, M_{u',\mu\nu}^{-1} \, M_{v,\nu\beta}^{-1/2} \; ,
\ee
where 
$M_{v,\alpha\beta}^{-1/2} =  \delta_{\alpha\beta} - v_\alpha v_\beta /(1+v^2+\sqrt{1+v^2})$,
and $u'_\alpha = \sum_\mu M_{v,\alpha\mu}^{-1/2} u_\mu$.

After some algebra, using Eq. (\ref{Ap:InvM}) we can express the inverse of the covariance,
Eq. (\ref{PixCovP1h}), as:
\bea
\label{Ap:InvCov}
C_{\alpha\beta}^{-1} (\vec{x},\vec{x}{\,}') 
& \rightarrow & \delta_D (\vec{x},\vec{x}{\,}') 
\\ \nonumber
& & \times \left[ \delta_{\alpha\beta} \bar{n}_\alpha -  \bar{n}_\alpha 
	\frac{
	 P^{2h}_{\alpha\beta} 
	+ P^{1h}_{\alpha\beta} 
	+ Y_{\alpha\beta}
	}{1 + {\cal{T}}} \bar{n}_\beta 
	\right] \; ,
\eea
where the cross-term is:
\bea
 Y_{\alpha\beta} &=& \sum_\mu  \bar{n}_\mu
 \left( P^{2h}_{\alpha\mu}  \, P^{1h}_{\mu\beta}
 + P^{1h}_{\alpha\mu}  \,  P^{2h}_{\mu\beta} \right.
 \\ \nonumber
 & & \left. - P^{2h}_{\alpha\beta}  \, P^{1h}_{\mu\mu}
  - P^{1h}_{\alpha\beta} \,  P^{2h}_{\mu\mu} \right) \; ,
\eea
and the term appearing of the denominator in Eq. (\ref{Ap:InvCov}) is:
\be
{\cal{T}} = \sum_\mu 
\bar{n}_\mu \, (P^{2h}_{\mu\mu} + P^{1h}_{\mu\mu}) +
\sum_{\mu\nu} \bar{n}_\mu  \bar{n}_\nu ( P^{2h}_{\mu\mu} \, P^{1h}_{\nu\nu}
- P^{2h}_{\mu\nu} \, P^{1h}_{\mu\nu} ) \; .
\ee
Compare this result with Eq. (\ref{Eq:InvCov}). We detect some familiar
expressions, in particular: 
\be
{\cal{P}} = \sum_\mu \bar{n}_\mu B_\mu^2 P_m 
\equiv \sum_\mu {\cal{P}}_\mu = \sum_\mu \bar{n}_\mu P^{2h}_{\mu\mu} \; .
\ee
It is now useful to rename the clustering strength of the 2-halo term 
as ${\cal{P}}_\mu \to {\cal{P}}^{2h}_\mu$, ${\cal{P}} \to {\cal{P}}^{2h}$,
and to define the 1-halo clustering strength as 
${\cal{P}}^{1h} = \sum_\mu \bar{n}_\mu P^{1h}_{\mu\mu} = \sum_\mu {\cal{P}}^{1h}_\mu $.
The cross-terms mixing the 1-halo and the 2-halo terms appear in the combinations:
\bea
\label{Ap:X}
P^c_{\alpha\beta} &\equiv& \sum_\mu P^{2h}_{\alpha\mu} \, \bar{n}_\mu \, P^{1h}_{\mu\beta} \; ,
\\ \nonumber
{\cal{P}}^c_{\alpha\beta} &\equiv& \bar{n}_\alpha \bar{n}_\beta P^{2h}_{\alpha\beta} P^{1h}_{\alpha\beta} \; .
\eea
Once again, we find it useful to 
define the dimensionless clustering strengths of these cross-terms, as was done
for the 2-halo and the 1-halo terms. They are:
\bea
{\cal{P}}^c_{\alpha} &\equiv& \bar{n}_\alpha P^c_{\alpha\alpha} 
= \sum_\beta {\cal{P}}^c_{\alpha\beta}  \; ,
\\ \nonumber
{\cal{P}}^c &\equiv& \sum_\alpha \bar{n}_\alpha P^c_{\alpha\alpha} 
= \sum_{\alpha\beta} {\cal{P}}^c_{\alpha\beta} \; .
\eea
With these definitions we find that:
\be
{\cal{T}} = {\cal{P}}^{2h} + {\cal{P}}^{1h} + {\cal{P}}^{2h} {\cal{P}}^{1h} -
{\cal{P}}^c \; .
\ee
Similarly, we get:
\be
 Y_{\alpha\beta} = P^{c}_{\alpha\beta} +  P^{c}_{\beta\alpha}
- P^{2h}_{\alpha\beta}  \, {\cal{P}}^{1h}
- P^{1h}_{\alpha\beta}  \, {\cal{P}}^{2h}
\; .
\ee

The Fisher matrix was defined in a generic sense in Eq. (\ref{Def:FC}). 
That definition, as well as the construction of the optimal quadratic estimators, 
are valid for any Gaussian variables \citep{1998ApJ...499..555T}. 
In a related result,
\citet{SmithMarian15} recently derived an optimal estimator for the matter
power spectrum, as well as the Fisher matrix for the power spectrum, 
including not only the 1-halo term, but also the 2- and 3-halo contributions 
to the trispectrum --- most of which are, strictly speaking, non-Gaussian contributions.
We did include the 1-halo term in the pixel covariance,  
as well as in the trispectrum, but only through the assumption of Gaussianity of the
4-point function. Due to the non-Gaussian terms that will appear in
the trispectrum, our estimators are not exactly optimal. 
Nevertheless, in some sense our result are more general than those of 
\citet{SmithMarian15}, since the multi-tracer estimators can be employed not 
only in the computation of the matter power spectrum, but also for the biases 
and the RSDs.

Since we keep the assumption of Gaussianity, 
all we have to do is work out the algebra with the covariance of Eq. (\ref{PixCovP1h}),
and its inverse, given by Eq. (\ref{Ap:InvCov}).
After a lengthy calculation, we find that the Fisher matrix which generalizes the
expression in the integrand of Eq. (\ref{Eq:FBin}) can be expressed as:
\bea
\nonumber
{\cal{F}}_{\mu\nu}^{2h} &=& \frac{1}{4 (1+{\cal{T}})^2} 
\left\{ \; \left[ (1+{\cal{P}}^{1h} ){\cal{P}}^{2h} - {\cal{P}}^c \right] 
\right.
\\ \nonumber
& & \times \left[ \delta_{\mu\nu} {\cal{P}}^{2h}_\mu (1+{\cal{T}}) 
	- (1+{\cal{P}}^{1h}) {\cal{P}}^{2h}_\mu {\cal{P}}^{2h}_\nu \right.
\\ \nonumber
& & \left. 	+ \; (1+{\cal{P}}^{2h}) {\cal{P}}^c_{\mu\nu} + {\cal{P}}^{2h}_\mu {\cal{P}}^c_\nu 
	+ {\cal{P}}^{2h}_\nu {\cal{P}}^c_\mu
	\right] 
\\ \nonumber
& & \, + \; (1+{\cal{P}}^{1h})^2 \, {\cal{P}}^{2h}_\mu {\cal{P}}^{2h}_\nu
	- ({\cal{P}}^c)^2 {\cal{P}}^c_{\mu\nu} 
\\ \label{Ap:Fisher}
& & \, - \left. (1+ {\cal{P}}^{1h}) \, ({\cal{P}}^{2h}_\mu {\cal{P}}^c_\nu 
+ {\cal{P}}^{2h}_\nu {\cal{P}}^c_\mu) \; \right\} \; .
\eea
It can be easily verified that taking $P^{1h}_{\mu\nu} \to 0$ implies ${\cal{P}}^c_{\mu\nu} \to 0$,
${\cal{T}} \to {\cal{P}}^{2h}$, etc., and this expression reduces to the matrix
${\cal{F}}_{\mu\nu}$ which is inside the integral in Eq. (\ref{Eq:FBin}).
The matrix is also manifestly symmetric.

It can also be shown that this Fisher matrix is positive-definite, with positive diagonal terms
and a positive determinant. This guarantees that the covariance of the 2-halo power 
spectrum is also positive-definite.

We should stress once again that the expression above is only valid in 
the approximation that the 1-halo term is separable, i.e., 
$P^{1h}_{\alpha\beta}/\sqrt{P^{1h}_{\alpha\alpha} P^{1h}_{\beta\beta} } 
\simeq \delta_{\alpha\beta}$.
For a general form of the 1-halo term, the pixel covariance matrix cannot be inverted 
analytically, which means that there are no closed-form expressions for the Fisher matrix or
for the optimal weights. One could still go ahead and compute them numerically, 
without any difficulty.


\subsection{Fisher matrix and optimal weights for the 1-halo term}

Now, suppose that what we are in fact interested in measuring the 1-halo term.
The 1-halo term is now the ``signal'', while the the 2-halo term, as well as shot 
noise, become the ``noise''. This should in fact be the case for very small scales 
($k \gg 1 \, h$/Mpc), where the 1-halo term dominates over the 
2-halo term \citep{CooraySheth}.

The pixel covariance is still the same, as in Eq. (\ref{PixCovP1h}),
and, as long as the 1-halo term is separable, the inverse covariance is also 
unaltered --- see Eq. (\ref{Ap:InvCov}).
The basic difference is that now instead of writing 
$P^{2h}_{\alpha\beta} = B_\alpha (z,k,\mu_k) B_\beta (z,k,\mu_k) P_m(k)$, we assume that
the 1-halo term can be written effectively as something like
$P^{1h}_{\alpha\beta} = H_\alpha (z,k,\mu_k) H_\beta (z,k,\mu_k) U(k)$, where $U(k)$
contains information about the shape of the mean halo profile.
This is a strong assumption: it means that, for $N$ species of tracers, the 
1-halo term would have only $N$ degrees of freedom (the $P^{1h}_\alpha = H_\alpha^2 U$), 
while the full expression in fact has $N(N-1)/2$ degrees of freedom.

Keeping the hypothesis that the 1-halo term is separable,
all we need to do now, in order to find its Fisher matrix, 
is to exchange all the 2-halo terms by the 1-halo terms in Eq. (\ref{Ap:Fisher}).
This procedure can also be used to define the optimal weights that ought 
to be used when extracting information about the 1-halo term from 
galaxy surveys.

Since many of the objects defined above are already symmetric under
the exchange $P^{2h}_{\mu\nu} \leftrightarrow P^{1h}_{\mu\nu}$
(this includes ${\cal{T}}$, $Y_{\mu\nu}$ and ${\cal{P}}^c_{\mu\nu}$), the
1-halo Fisher matrix can be immediately written as:
\bea
\nonumber
{\cal{F}}_{\mu\nu}^{1h} &=& \frac{1}{4 (1+{\cal{T}})^2} 
\left\{ \; \left[ (1+{\cal{P}}^{2h} ){\cal{P}}^{1h} - {\cal{P}}^c \right] 
\right.
\\ \nonumber
& & \times \left[ \delta_{\mu\nu} {\cal{P}}^{1h}_\mu (1+{\cal{T}}) 
	- (1+{\cal{P}}^{2h}) {\cal{P}}^{1h}_\mu {\cal{P}}^{1h}_\nu \right.
\\ \nonumber
& & \left. 	+  \; (1+{\cal{P}}^{1h}) {\cal{P}}^c_{\mu\nu} + {\cal{P}}^{1h}_\mu {\cal{P}}^c_\nu 
	+ {\cal{P}}^{1h}_\nu {\cal{P}}^c_\mu
	\right] 
\\ \nonumber
& & \, + \; (1+{\cal{P}}^{2h})^2 \, {\cal{P}}^{1h}_\mu {\cal{P}}^{1h}_\nu
	- ({\cal{P}}^c)^2 {\cal{P}}^c_{\mu\nu} 
\\ \label{Ap:Fisher1h}
& & \, - \left. (1+ {\cal{P}}^{2h}) \, ({\cal{P}}^{1h}_\mu {\cal{P}}^c_\nu 
+ {\cal{P}}^{1h}_\nu {\cal{P}}^c_\mu) \; \right\} \; .
\eea
For very small scales the 2-halo term can be neglected, and we are left just with
the 1-halo terms.

The Fisher matrix in bins of $k$ is just as in Eq. (\ref{Eq:FBin}):
\be
\label{Ap:FBin1h}
F_{\mu,i \, ; \, \nu,j}^{1h} = \frac{\delta_{ij}}{(P^{1h}_{\mu\nu})^2}  
\int_{V_i} \frac{d^3 x \, d^3 k}{(2\pi)^3}
{\cal{F}}_{\mu\nu}^{1h} \; .
\ee
The optimal weights follow in a straightforward manner from this expression, just
as was done for the 2-halo term.


\subsection{Joint Fisher matrix for the 2-halo and 1-halo terms}

The next obvious question is: what if we wish to estimate {\em both} the 2-halo and the
1-halo terms in a multi-tracer cosmological survey, simultaneously? 
The two contributions are clearly correlated, so their information contents are not
independent. Evidently, on either very large or very small scales the correlations
between the two are small, and one can treat the signal 
($P^{2h}$ on large scales; $P^{1h}$ on small scales) as effectively
independent of the noise. However, on intermediate scales (around $k\sim 1 \, h$/Mpc)
the 1-halo and the 2-halo terms may have significant correlations.
Furthermore, the approximation of separable 1-halo term becomes more
accurate on those intermediate scales.

The pixel covariance is still given by Eq. (\ref{PixCovP1h}), 
and its inverse also remains unchanged, but 
we would now be considering our ``signal'' as the sum 
$ P^t_{\mu} \equiv P^{2h}_{\mu} + P^{1h}_{\mu} $.
The main difference is that the derivatives of the pixel covariance,
which in the case when we neglected the 1-halo term were given by Eq. (\ref{Eq:dCdP}), 
should now be computed with respect to this total contribution, i.e.:
\bea
\nonumber
\frac{\partial C_{\alpha\beta} (\vec{x},\vec{x}{ \, }')}{\partial P^t_{\mu,i}}
& = & 
\frac{\partial P^{2h}_{\mu,i}}{\partial P^{t}_{\mu,i}}
\frac{\partial C_{\alpha\beta} (\vec{x},\vec{x}{ \, }')}{\partial P^{2h}_{\mu,i}}
+
\frac{\partial P^{1h}_{\mu,i}}{\partial P^{t}_{\mu,i}}
\frac{\partial C_{\alpha\beta} (\vec{x},\vec{x}{ \, }')}{\partial P^{1h}_{\mu,i}}
\\ \nonumber
&=& \int \frac{d^3k }{(2\pi)^3} 
e^{i \vec{k} \cdot (\vec{x}-\vec{x}{ \, }')} 
\left( \delta_{\alpha\mu} \delta^i_{\vec{x},\vec{k}} +
\delta_{\beta\mu} \delta^i_{\vec{x}{ \, }',\vec{k}} \right)
\\ \nonumber
& & 
\times \left[ 
\frac{B_\alpha (\vec{x},\vec{k}) B_\beta (\vec{x}{ \, }',\vec{k})}{2 \, B_{\mu}^2(\vec{x}_i,\vec{k}_i)} 
\right.
\\ \label{Eq:dCdPt}
& & 
+ \left.
\frac{H_\alpha (\vec{x},\vec{k}) H_\beta (\vec{x}{ \, }',\vec{k})}{2 \, H_{\mu}^2(\vec{x}_i,\vec{k}_i)} 
\right]
\; .
\eea
Substitution of this expression, together with the inverse of the pixel covariance, into
Eq.(\ref{Def:FC}), leads to the full Fisher matrix for the power spectrum.
This can be written as:
\bea
\label{FullFish}
F^t_{\mu,i;\nu,j} &=& \delta_{ij} 
\int_{V_i} \frac{d^3 x \, d^3 k}{(2\pi)^3}
\left[ \frac{{\cal{F}}^{2h}_{\mu\nu} }{(P^{2h}_{\mu\nu,i})^2}  
+
\frac{{\cal{F}}^{1h}_{\mu\nu} }{(P^{1h}_{\mu\nu,i})^2}  
\right.
\\ \nonumber
& & 
+ \left.
\left( \frac{1}{P^{2h}_{\mu\mu,i} P^{1h}_{\nu\nu,i}}  
+ \frac{1}{P^{2h}_{\nu\nu,i} P^{1h}_{\mu\mu,i}}  \right)
{\cal{F}}^c_{\mu\nu} \right] \; ,
\eea
where  ${\cal{F}}^c_{\alpha\beta}$ contains the cross-terms between the 1-halo and the 2-halo 
terms which follow from Eq. (\ref{Eq:dCdPt}).
Expressing the inverse covariance of Eq. (\ref{Ap:InvCov}) in terms of 
$C^{-1}_{\alpha\beta} (\vec{x},\vec{x}{\,}') = \delta_D (\vec{x},\vec{x}{\,}') D_{\alpha\beta}$,
the information mixing between the 1-halo and 2-halo terms is given by:
\bea
\nonumber
{\cal{F}}^c_{\mu\nu} &=& \frac{1}{8}
\sum_{\alpha\beta} \left( 
P^{1h}_{\alpha\beta} D_{\alpha\beta}  \,
P^{2h}_{\mu\nu} D_{\mu\nu} 
+ 
P^{2h}_{\alpha\beta} D_{\alpha\beta}  \,
P^{1h}_{\mu\nu} D_{\mu\nu} \right.
\\ \label{FMixed}
& & + \left.
P^{2h}_{\nu\alpha}
D_{\alpha\mu} 
P^{1h}_{\mu\beta}
D_{\beta\nu}
+
P^{1h}_{\nu\alpha}
D_{\alpha\mu} 
P^{2h}_{\mu\beta}
D_{\beta\nu}
\right) \; .
\eea
It is trivial to obtain the full expression, although it turns out to be rather long.
It can be significantly simplified if we employ two additional auxiliary definitions:
\bea
\label{Ap:Z}
Z^{1,2h}_{\mu\nu} &\equiv& \sum_\alpha P^{1,2h}_{\mu\alpha} D_{\alpha\nu} \; ,
\\ \label{Ap:W}
W^{1,2h}_{\mu\nu} &\equiv& P^{1,2h}_{\mu\nu} D_{\mu\nu} \; .
\eea
In terms of these variables we have, e.g.:
\be
\label{Ap:Ze}
Z^{2h}_{\mu\nu} = \frac{\bar{n}_\nu}{1+{\cal{T}} }
\left[ \left( 1 + 2 {\cal{P}}^{2h} + {\cal{P}}^{1h} \right) P^{2h}_{\mu\nu} + P^c_{\mu\nu} \right] \; ,
\ee
and
\bea
\label{Ap:We}
W^{2h}_{\mu\nu} &=& \delta_{\mu\nu} {\cal{P}}^{2h}_\mu
+ 	\frac{
	\left( 1 - {\cal{P}}^{1h} \right) {\cal{P}}^{2h}_{\mu}{\cal{P}}^{2h}_{\nu}
		}{1+{\cal{T}} }
\\ \nonumber
& & + 	\frac{
		{\cal{P}}^{2h}_{\mu}{\cal{P}}^{c}_{\nu}
		+ {\cal{P}}^{2h}_{\nu}{\cal{P}}^{c}_{\mu}
		+ \left( 1 - {\cal{P}}^{2h} \right) {\cal{P}}^{c}_{\mu\nu}
		}{1+{\cal{T}} }
\; ,
\eea
as well as the analogous expressions obtained by exchanging $1h \leftrightarrow 2h$.
In terms of these definitions we have:
\bea
\nonumber
{\cal{F}}^c_{\mu\nu} &=& \frac{1}{8}
\left[ Z^{1h} W^{2h}_{\mu\nu} +
Z^{2h} W^{1h}_{\mu\nu} \right.
\\ \nonumber
& & 
+ \left. \sum_\alpha \left( 
Z^{2h}_{\mu\alpha} Z^{1h}_{\alpha\nu}
+ Z^{1h}_{\mu\alpha} Z^{2h}_{\alpha\nu} \right) \right]
\; ,
\eea
where $Z=\sum_\mu Z_{\mu\mu} = \sum_{\mu\nu} W_{\mu\nu}$.
In fact, these definitions are also helpful when computing ${\cal{F}}^{2h}_{\mu\nu}$ 
(taking all $Z \to  Z^{2h}$ and $W \to  W^{2h}$) and
${\cal{F}}^{1h}_{\mu\nu}$ (taking all $Z \to  Z^{1h}$ and $W \to  W^{1h}$).


\section{Conclusions}
\label{S:Conclusions}

We have obtained optimal estimators for the Fourier analysis of multi-tracer
cosmological surveys. The formulas were derived in Sec. 3, and a practical
algorithm for the Fourier analysis of multi-tracer surveys 
was summarized in Sec. \ref{SubSec:Algo}. Those are
the main results of this paper.

The multi-tracer technique estimates the individual redshift-space power spectra for
each tracer, $P_\alpha (z,k,\mu_k)$, taking into account the covariance between
the tracers which is induced by the large-scale structure. 
In contrast to the estimators obtained by
\citet{Percival:2003pi} or \citet{SmithMarian15}, which are suited for
estimating the underlying matter power spectrum {\em after} fixing the biases
and the RSDs, our optimal estimators can be used to measure both 
the power spectrum, the biases, the shape of RSDs, etc.
In particular, our estimators facilitate measurements of RSDs, scale-dependent 
bias and non-Gaussianities from cosmological surveys of 
multiple tracers, helping realize the potential for determining
those physical parameters to an
accuracy which is not limited by cosmic variance
\citep{SeljakNG,McDonald:2008sh,GM2010,Hamaus2011,AL13}.

We also included the contribution from the 1-halo term in our calculations 
(Sec. \ref{S:1-halo}).
Although on very large scales ($k \ll 1 \, h$ Mpc$^{-1}$) the 2-halo term is dominant, 
the 1-halo term gives a nearly-constant contribution in that limit, adding to shot noise 
--- and, unlike shot noise, it does affect the cross-correlations.

It is important to stress that
our formulas are relatively simple generalizations of those by FKP 
\citep{1994ApJ...426...23F} and PVP \citep{Percival:2003pi}, so readers 
familiar with these standard methods should have no trouble implementing the
multi-tracer technique. We tested the estimators (see Sec. \ref{S:Applications}) in a 
wide variety of situations, and they performed quite robustly --- in many instances, 
significantly better than the FKP method. It should now be 
straightforward to combine cosmological surveys targeting different
types of galaxies, quasars and other tracers of large-scale structure.


\vskip 0.5cm

\noindent {\it  Acknowledgements} -- 
We would like to thank FAPESP, CNPq, and
the University of S\~ao Paulo's {\em NAP LabCosmos} for financial support.


\appendix
\section{Degenerate tracers in the basis of the auto-power spectra}

Suppose we have $N$ types of tracers, 
but we would like to combine the last 2 of those 
tracers into a {\em single} species, for a new total of $N-1$ tracers.
For simplicity, let's regard our original parameters as 
${\cal{P}}_\mu$ ($\mu = 1 \ldots N$). 
We would like to change variables to ${\cal{P}}'_a$ ($a = 1 \ldots N-1$), where  
${\cal{P}}'_a= {\cal{P}}_a$ for $a = 1, \ldots N-2$, and the new tracer 
species is constructed by combining the last two tracers,
${\cal{P}}'_{N-1} = {\cal{P}}_{N-1} + {\cal{P}}_{N}$.
When the biases of the two species which are combined into one are identical, 
$B_{N-1}=B_{N-2}=B'_{N-1}$, this linear combination ensures that 
${\cal{P}}'_{N-1} = (\bar{n}_{N-1} + \bar{n}_{N}) B'^2_{N-1} P_m$, so
clearly $\bar{n}'_{N-1} = \bar{n}_{N-1} + \bar{n}_{N}$ --- i.e., the total number of 
galaxies of the new species is the sum of the number of galaxies of the
two original species.

The Jacobian for the transformation ${\cal{P}}'_a \to {\cal{P}}_\mu$
is $\partial {\cal{P}}'_a/\partial {\cal{P}}_\mu$, and this Jacobian
equals $\delta_{a\mu}$ when $\mu < N-1$, 
it vanishes if $a = N-1$ and $\mu < N-1$, and
it is equal to 1 if $a = N-1$ and $\mu \geq N-1$.

However, what we need for the new Fisher matrix is the Jacobian for the 
inverse transformation\footnote{In fact, since this Jacobian is not a 
square matrix, it only has a pseudo-inverse. However, in this case the 
pseudo-inverse is exact.}, 
${\cal{P}}_\mu \to {\cal{P}}'_a$, i.e., 
$J_{\mu a} = (\partial {\cal{P}}_\mu/\partial {\cal{P}}'_a)^{-1}$.
But this turns out to be a very simple matrix: 
$J_{\mu a} = \delta_{\mu a}$ when $\mu < N-1$, 
it vanishes if $a = N-1$ and $\mu < N-1$, and
when $a = N-1$ and $\mu \geq N-1$ the Jacobian is equal to 
${\cal{P}}_\mu/{\cal{P}}'_{N-1} = 
{\cal{P}}_\mu/({\cal{P}}_{N-1}+{\cal{P}}_{N-1})$
 --- see also the 
discussion at the beginning of Sec. \ref{Sec:Prop}.

Hence, in the new variables the Fisher matrix (or, more precisely, 
the Fisher information density per unit of phase space volume) is:
\be
F'_{ab}  =  F[{\cal{P}}'_a,{\cal{P}}'_b] = \sum_{\mu\nu}^N 
J_{\mu a} \,
\frac{{\cal{F}}_{\mu\nu} }{{\cal{P}}_\mu \, {\cal{P}}_\nu} \,
J_{\nu b} \; ,
\ee
where ${\cal{F}}_{\mu\nu}  = F[\log {\cal{P}}_\mu,\log {\cal{P}}_\nu]$ 
--- see Eq. (\ref{Def:Fcurved}).
This turns out to be given by:
\be
\label{Eq:FNEWSUM}
F'_{ab}
 = 
\left( 
\begin{array}{c;{2pt/2pt}c}
\frac{1}{{\cal{P}}'_a \, {\cal{P}}'_b} {\cal{F}}_{ab}
& \frac{1}{{\cal{P}}'_a {\cal{P}}'_{N-1}} 
\sum_{\nu=N-1}^N  {\cal{F}}_{a\nu} \\
\; & \; \\
\hdashline[2pt/2pt] \\
{\rm Sym} &
\frac{1}{({\cal{P}}'_{N-1})^2} 
\sum_{\mu,\nu=N-1}^N 
{\cal{F}}_{\mu \nu}
\end{array}
\right) \; ,
\ee
where the upper left block is an $(N-2) \times (N-2)$ matrix, the right block is an
$1 \times (N-2)$ column, the lower left block is an $(N-2) \times 1$ row, and the
lower right block is a single entry.
Hence, the resulting $(N-1)$-dimensional Fisher matrix is 
given simply by summing the lines and columns corresponding to the two tracers
which were combined into a single type. 

Now, it can be easily verified from Eq. (\ref{Def:Fcurved}) that summing any two 
lines and columns of the fisher matrix ${\cal{F}}_{\mu\nu}$ 
yields precisely the Fisher matrix where the new entries correspond to
the Fisher information for the {\em sum} of the clustering strengths of the two
species that were combined. In other words, if we take Eq. (\ref{Def:Fcurved})
and use ${\cal{P}}'_a$ to compute ${\cal{F}}'_{ab} = F[\log {\cal{P}}'_a , \log {\cal{P}}'_b]$,
then the Fisher matrix $F'_{ab} = {\cal{F}}'_{ab}/{\cal{P}}'_a {\cal{P}}'_b$
is identical to Eq. (\ref{Eq:FNEWSUM}).

This argument can be iterated to show that combining
any number of tracers into a single species corresponds to adding their 
clustering strengths, and this operation results in a simple sum of the Fisher 
information of those tracers.

\section{Inversion of the covariance matrix}

Consider a matrix of the form  $M_{v} = \mathbb{1} \, + \, v \otimes v$
--- i.e., $M_{v, \, \mu\nu} = \delta_{\mu\nu} \, + \, v_\mu v_\nu$.
As discussed in Section 2, it can be shown that:
\be
\label{Ap:InvMf}
M^{-1}_{v} = \mathbb{1} - \frac{v \otimes v}{1+v^2} \; , 
\ee
where $v^2 = {\rm Tr} (v \otimes v) = \sum_\alpha v_\alpha v_\alpha $.
This is in fact a special case of the Sherman-Morrison-Woodbury formula
\citep{SMW}.

The matrix $M_v$ also has a simple ``square root'', as well as an 
``inverse square root'', given by:
\bea
\label{Ap:Sqrt}
M^{1/2}_{v} &=& \mathbb{1} + \frac{v \otimes v}{1+ \sqrt{1+v^2}} \; 
\\ 
\label{Ap:ISqrt}
M^{-1/2}_{v} &=& \mathbb{1} - \frac{v \otimes v}{1+v^2 + \sqrt{1+v^2}} \; , 
\eea
where $M_v^{-1/2} \cdot M_v^{-1/2} = M_v^{-1}$ and $M_v^{-1/2} \cdot M_v = M_v^{1/2}$,
from which follows that:
\be
M_v^{-1/2} \cdot M_v \cdot M_v^{-1/2} = \mathbb{1} \; .
\ee

Now, take a matrix $M = M_v + u \otimes u$. The first piece of that matrix can be 
diagonalized following the procedure outlined above, so we have that:
\bea
\nonumber
M_v^{-1/2} \cdot M \cdot M_v^{-1/2} 
& = & \mathbb{1} + (M_v^{-1/2} \cdot u) \otimes (u \cdot M_v^{-1/2} ) 
\\ \label{Ap:uprime}
& = & \mathbb{1} +  u' \otimes u' \; ,
\eea
where $u' =  M_v^{-1/2} \cdot u$ (i.e., $u'_\alpha =  \sum_\mu M^{-1/2}_{v, \alpha\mu} u_\mu$).
But the matrix of Eq. (\ref{Ap:uprime}) can now be inverted using the equivalent of 
Eq. (\ref{Ap:InvMf}), and moreover it has an inverse square root 
$M_{u'}^{-1/2}$, as in Eq. (\ref{Ap:ISqrt}).
Therefore, we have that:
\be
\nonumber
M_{u'}^{-1/2} \cdot M_v^{-1/2} \cdot M \cdot M_v^{-1/2} \cdot M_{u'}^{-1/2}
 =  \mathbb{1} \; .
\ee
Therefore, the inverse of the matrix $M$ is given by:
\bea
\nonumber
M^{-1} &=& M_v^{-1/2} \cdot M_{u'}^{-1/2} \cdot M_{u'}^{-1/2} \cdot M_v^{-1/2}
\\ \label{Ap:InvM2}
& = & M_v^{-1/2} \cdot M_{u'}^{-1} \cdot M_v^{-1/2} \; .
\eea
Of course, one could equally write this inverse as:
\be
\nonumber
M^{-1}  =  M_u^{-1/2} \cdot M_{v'}^{-1} \cdot M_u^{-1/2} \; ,
\ee
where $v' =  M_u^{-1/2} \cdot v$.



\bibliographystyle{mn2e.bst}

\end{document}